\documentclass[pra,twocolumn,superscriptaddress,amsmath,amssymb,floatfi,noshowpacs]{revtex4}
\usepackage{graphicx}

\begin{document}
\title{\textbf{Highly nonlocal optical nonlinearities in atoms trapped near a waveguide}}
\author{Ephraim Shahmoon}
\affiliation{Department of Physics, Harvard University, Cambridge MA 2138, USA}
\affiliation{Department of Chemical Physics, Weizmann Institute of Science, Rehovot, 7610001, Israel}
\affiliation{Department of Physics of Complex Systems, Weizmann Institute of Science, Rehovot, 7610001, Israel}
\author{Pjotrs Gri\v{s}ins}
\affiliation{Vienna Center for Quantum Science and Technology, Atominstitut, TU Wien, 1020 Vienna, Austria}
\author{Hans Peter Stimming}
\affiliation{Fakult{\"a}t f{\"u}r Mathematik, Universit{\"a}t Wien, 1090 Vienna, Austria}
\affiliation{Wolfgang Pauli Institute, Universit\"{a}t Wien, 1090 Vienna, Austria}
\author{Igor Mazets}
\affiliation{Vienna Center for Quantum Science and Technology, Atominstitut, TU Wien, 1020 Vienna, Austria}
\affiliation{Wolfgang Pauli Institute, Universit\"{a}t Wien, 1090 Vienna, Austria}
\affiliation{Ioffe Physico-Technical Institute of the Russian Academy of Sciences, 194021 St. Petersburg, Russia}
\author{Gershon Kurizki}
\affiliation{Department of Chemical Physics, Weizmann Institute of Science, Rehovot, 7610001, Israel}
\date{\today}

\begin{abstract}
Nonlinear optical phenomena are typically local. Here we predict the possibility of highly nonlocal optical nonlinearities for light propagating in atomic media trapped near a nano-waveguide, where long-range interactions between the atoms can be tailored. When the atoms are in an electromagnetically-induced transparency configuration, the atomic interactions are translated to long-range interactions between photons and thus to highly nonlocal optical nonlinearities. We derive and analyze the governing nonlinear propagation equation, finding a roton-like excitation spectrum for light and the emergence of order in its output intensity.
These predictions open the door to studies of unexplored wave dynamics and many-body physics with highly-nonlocal interactions of optical fields in one dimension.
\end{abstract}

 \maketitle

\section{Introduction}
Optical nonlinearities are commonly described by local nonlinear response of the material to the optical field, resulting in the dependence of the refractive index at point $z$ on the field at the same point, $E(z)$ \cite{BOYD}. Recently however there has been a growing interest in nonlocal nonlinear optics, namely, in mechanisms whereby the refractive index at $z$ depends on the field intensity at different points $z'$ in the material \cite{KRO}. Mechanisms that give rise to such nonlocal nonlinearities include e.g. heat diffusion \cite{HD1,HD2}, molecular reorientation in liquid crystals \cite{LC} and atomic diffusion \cite{AD1,AD2}. This paper discusses a new regime of \emph{extremely nonlocal} nonlinearities affecting both the frequency and the \emph{quadratic dispersion} of optical waves [see Eq. (\ref{NLSE}) below]. The physical mechanism that leads to this new regime is very different than those explored previously \cite{KRO,HD1,HD2,LC,AD1,AD2}: it relies on an atomic medium prepared in an electromagnetically-induced transparency (EIT) configuration, whose optical nonlinearity is \emph{controlled} by shaping the nonlocal \emph{dipolar interactions} between the atoms.

EIT is associated with the lossless and slow propagation of light pulses in resonant atomic medium subject to coherent driving of an auxiliary atomic transition \cite{FIM}. Since the early days of EIT, it has been explored as a means of enhancing optical nonlinearities \cite{FIM,SIM,HY,LI}.
A particulary effective mechanism for giant optical nonlinearities is provided by dipolar interactions between atoms that form the medium: Since EIT can be described by the propagation of the so-called dark-state polariton \cite{FL}, which is a superposition of the light field and an atomic spin wave, the inherently nonlocal dipolar interactions between atoms  are translated to nonlocal nonlinearities in polariton propagation. In the case of dipolar interactions between Rydberg atoms in free space, most theoretical \cite{FRD,EFI,GOR,HE1,PF,OTT,HE2,GRA,GOR2} and experimental \cite{OFER1,OFER2,ADM1,REM,HOF} studies have focused on their remarkable strength, a useful feature in quantum information, whereas their nonlocal aspect has received less attention \cite{POH}.

The present work rests on two recently explored mechanisms: that of EIT polaritons and that of \emph{modified long-range dipolar interactions} in confining geometries, such as fibers, waveguides, photonic band structures or transmission lines, which currently attracts considerable interest \cite{RDDI,vdWTL,KIM3,LIDDIna,CHA,CHA2,KUR,WAL,WAL2,KUR2}. Yet, we show that the combined effect of these mechanisms may allow for new and unfamiliar possibilities of nonlocal nonlinear optics. More specifically, we show how dispersive laser-induced dipolar interactions between atoms coupled to a nano-waveguide with a grating, which can be designed to extend over \emph{hundreds} of optical wavelengths \cite{LIDDIna,CHA}, are translated via EIT into extremely nonlocal optical nonlinearities. We then analyze light propagation in this medium along the waveguide and find a roton-like excitation spectrum for light and the emergence of spatial self-order in its output intensity.

\section{Summary and scope of the results}

The main general result of this work is the derivation of a nonlinear propagation equation for the (possibly quantum) EIT polariton field $\hat{\Psi}$ comprised of light \emph{guided} by the waveguide, and an atomic medium \emph{tightly} trapped along the waveguide axis $z$ with an effectively 1d inter-atomic potential $U(z-z')$,
\begin{eqnarray}
\left(\partial_t+v\partial_z\right)\hat{\Psi}(z)&=&-i\hat{\Delta}_c\alpha\hat{\Psi}(z)-i\hat{\Delta}_cCv^2\partial_z^2\hat{\Psi}(z),
\nonumber \\
\hat{\Delta}_c&=&\delta_c+\hat{\delta}_{NL},
\nonumber \\
\hat{\delta}_{NL}&=&\alpha\int_L dz'U(z-z')\hat{\Psi}^{\dag}(z')\hat{\Psi}(z').
\label{NLSE}
\end{eqnarray}
The left-hand side of the equation describes an envelope of a wave travelling with a group velocity $v$, whereas the first and second terms in the right-hand side present its frequency shift $\hat{\Delta}_c$ (multiplied by a coefficient $\alpha$) and its quadratic dispersion with a coefficient proportional to the detuning $\hat{\Delta}_c$ (and to a constant $C$), respectively. Nonlinearity comes about by noting the detuning $\hat{\Delta}_c$: it contains a linear component $\delta_c$ which is controlled by the EIT configuration (the so-called coupling-field detuning, see Fig. 1b), and a \emph{nonlocal} nonlinear detuning $\hat{\delta}_{NL}$ which depends on field intensities integrated over the medium with the interaction kernel $U(z)$. The appearance of a \emph{nonlocal nonlinearity} not only in the frequency shift but also in the \emph{dispersion coefficient} gives rise to a new regime of nonlocal nonlinear optics.
The physical system and reasoning that lead to Eq. (\ref{NLSE}), as well as its derivation, are discussed in Sec. III below. In principle, we assume sufficient conditions for a lossless EIT propagation (coefficients $v$, $\alpha$ and $C$ are real), whereas loss and decoherence mechanisms due to imperfections and scattering are analyzed in Sec. VI.

\begin{figure}
\begin{center}
\includegraphics[scale=0.30]{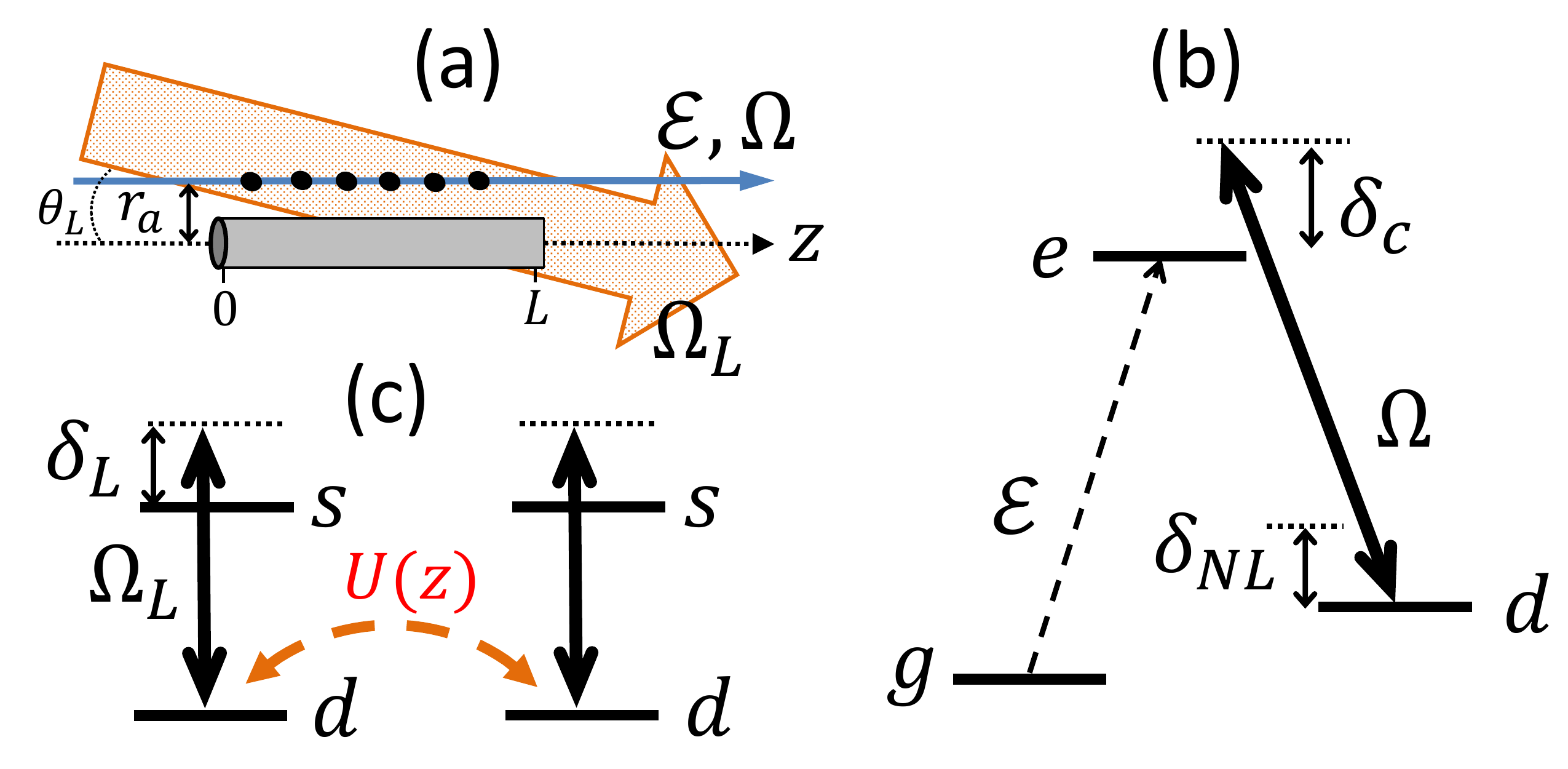}
\caption{\small{
(a) Setup: atoms (black dots), illuminated by the EIT fields [see (b)] $\hat{\mathcal{E}}$ and $\Omega$ (thin blue arrow), are trapped at a distance $r_a$ from a nano-waveguide (gray cylinder) along $z$, from $z=0$ to $z=L$. A far-detuned laser $\Omega_L$ (thick orange arrow), tilted by an angle $\theta_L$ from the $z$ axis, induces long-range interactions between the atoms, mediated by the waveguide modes [see (c)].
(b) EIT atomic configuration: the probe field $\hat{\mathcal{E}}$ is resonantly coupled to the $|g\rangle\rightarrow |e\rangle$ transition whereas the coupling field $\Omega$ is coupled to the $|d\rangle\rightarrow |e\rangle$ transition with detuning $\delta_c$. Interaction between the atoms in the $|d\rangle$-level [see (c)] induces its energy shift $\delta_{NL}$ which is effectively added to the detuning $\delta_c$. (c) Laser-induced dipolar interactions: the laser with Rabi frequency $\Omega_L$ and detuning $\delta_L$ operates on the $|d\rangle\rightarrow |s\rangle$ transition of all atoms, $|s\rangle$ being an additional level, thus inducing a dipolar potential $U(z)$ between pairs of atoms ($z$ apart) populating the state $|d\rangle$ \cite{LIDDI}.
}}
\end{center}
\end{figure}

The second part of this work (Sec. IV and V) is dedicated to study some first implications on wave propagation in such a medium, and specifically to the analysis of wave excitations on top of a continuous wave (CW) background. General results for the excitation spectrum (dispersion relation) and the intensity correlations at the output are presented in Sec. IV, Eqs. (\ref{bog}) and (\ref{g2}), respectively. These are followed in Sec. V by specific results for a medium of atoms trapped near a waveguide-grating which supports extremely long-range interactions $U(z)$, as per Eq. (\ref{UL}). The results for the excitation spectrum exhibit roton-like narrow-band shape, which may be probed by a homodyne detection scheme (Fig. 3). The roton-like spectrum signifies the tendency of light in this regime to exhibit spatial self-order, namely, crystal-like correlations; these can be revealed by measuring the photon intensity at the waveguide's output (Fig. 4).

The predicted self-order of light constitutes a new, hitherto unexplored, optical "phase", analogous to the spatial structure of cold atomic media subject to light-induced dipolar interactions \cite{LIDDIna,GIO,OD,ESS,CHA1,RIT2,RIT}. We discuss important aspects and prospects of this work in Sec. VII.

\section{EIT polaritons with nonlocal interactions}
\subsection{The system}
Consider a medium of identical atoms in an EIT configuration as in Fig. 1: The atoms are trapped at a distance $r_a$ from a nano-waveguide along its longitudinal $z$ axis \cite{RAU1,KIM3,RAU2,KIM1,KIM2,LUK,LOD,OROZ,WALM,GAE} (Fig. 1a). A strong (external) coupling field with a constant and uniform Rabi frequency $\Omega$ drives the $|d\rangle\rightarrow |e\rangle$ atomic transition with detuning $\delta_c$ and wavenumber $k_c$, whereas a weak (possibly quantized) probe field with carrier frequency $\omega_0$, wavenumber $k_0$ and envelope $\hat{\mathcal{E}}$ is resonantly coupled to the $|g\rangle\rightarrow |e\rangle$ transition ($\omega_0=\omega_{eg}$) [Fig. 1b]. Under tight transverse trapping (around $r_a$) with respect to the transition wavelength, the atomic positions can be characterized by their longitudinal component $z$ (Appendix A1) \cite{RAU1,LIDDIna}. We assume the existence of a dipolar interaction $U(z)$ between atoms that occupy the state $|d\rangle$ (Fig. 1c and Sec. V A below).
Then, the energy of level $|d\rangle$ of an atom at $z$ is shifted by $\delta_{NL} \sim n_a\int dz' U(z-z') P_{d}(z')$, $n_{a}$ being the atomic density (per unit length) and $P_{d}(z')$ the occupation of state $|d\rangle$ in an atom at $z'$.

\begin{figure}
\begin{center}
\includegraphics[scale=0.32]{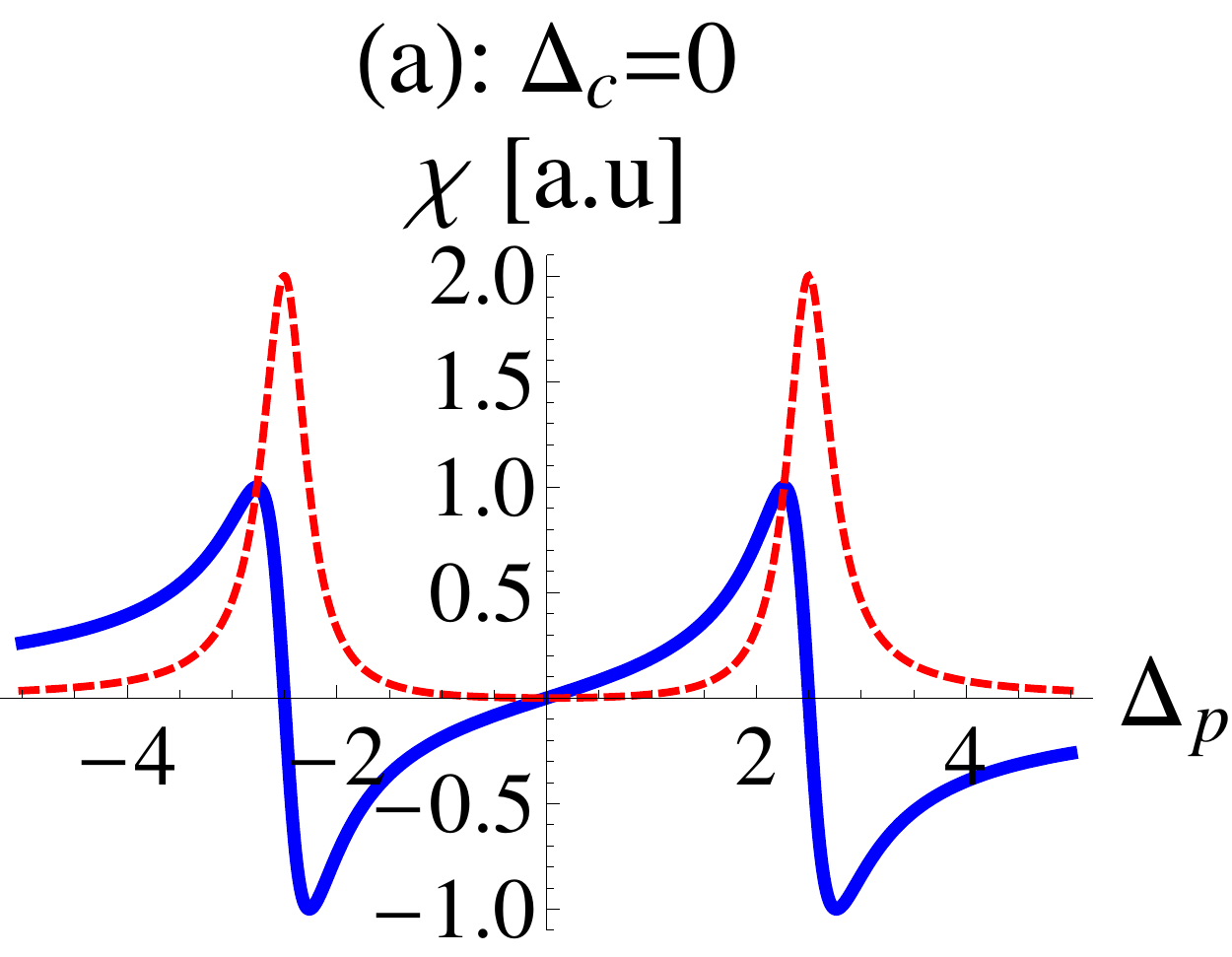}
\includegraphics[scale=0.32]{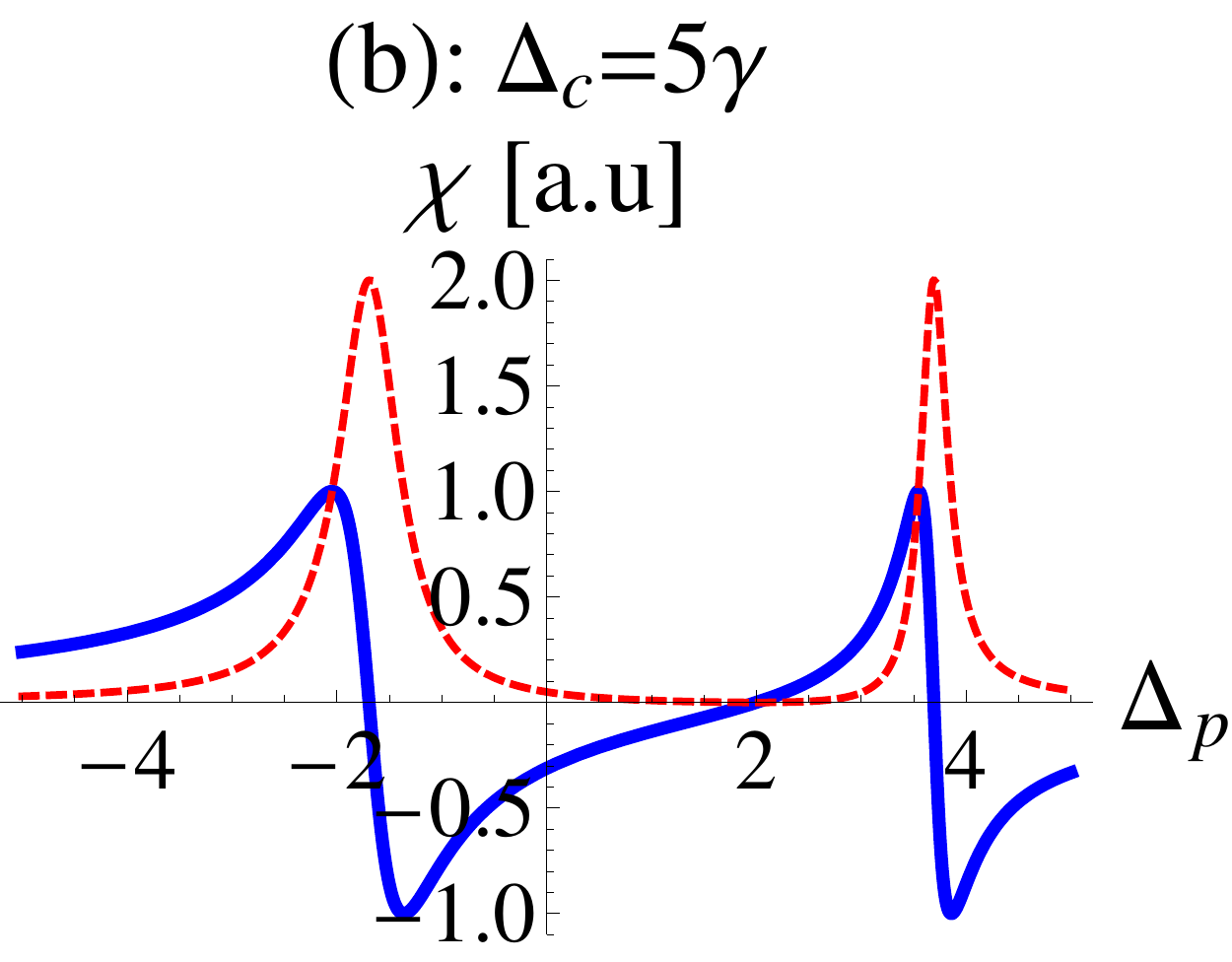}
\caption{\small{
Linear susceptibility of the EIT atomic medium to the probe field as a function of its detuning $\Delta_p$ \cite{FIM}. (a) For total detuning of the coupling field $\Delta_c=0$, the absorption $\mathrm{Im}\chi$ (red dashed line) and dispersion $\mathrm{Re}\chi$ (blue solid line) are symmetric and antisymmetric, respectively, with respect to $\Delta_p$, so that no (real) quadratic dispersion exists for the probe envelope centered around $\Delta_p=0$. (b) For $\Delta_c\neq0$, $\mathrm{Re}\chi$ is not antisymmetric, so that quadratic dispersion exists, giving rise to the term $\Delta_c Cv^2\partial_z^2$ in Eq. (\ref{NLSE}), with $\Delta_c=\delta_c+\delta_{NL}$ (Fig. 1b).
}}
\end{center}
\end{figure}
We may now explain the physical reasoning that leads to Eq. (\ref{NLSE}). In Fig. 2 we plot the complex linear susceptibility $\chi$ of the EIT medium to the probe field $\hat{\mathcal{E}}$ as a function of its detuning $\Delta_p$, in the presence of a coupling field detuned by $\Delta_c$ \cite{FIM}, which in our case is given by $\Delta_c=\delta_c+\delta_{NL}$ (Fig. 1b). When $\Delta_c=0$ (Fig. 2a), the absorption coefficient $\mathrm{Im}\chi$ is symmetric with respect to $\Delta_p$ whereas the dispersion $\mathrm{Re}\chi$ is antisymmetric, so that no (real) quadratic dispersion exists for the probe envelope centered around $\Delta_p=0$. By contrast, when $\Delta_c\neq0$ is introduced (Fig. 2b), $\mathrm{Re}\chi$ is no longer antisymmetric and quadratic dispersion exists, which explains the term $\Delta_c Cv^2\partial_z^2$ in Eq. (\ref{NLSE}). However, this comes at the price of non-vanishing losses at $\Delta_p=0$. For this reason we choose to work in the so-called Autler-Townes regime, $\Omega\gg \gamma,\Delta_c$ \cite{FIM}, where $\gamma$ is the width of the level $|e\rangle$. Then, for $\Delta_c$ smaller than the single-atom transparency window, $\Delta_c\ll \Omega^2/\gamma,\Omega$, but still larger than $\gamma$, the absorption per atom can become negligible while dispersion is still significant, as illustrated in Fig. 2b (see also Appendix A). This explains the lossless propagation described by Eq. (\ref{NLSE}) with real parameters $\alpha,v,C$. As long as the absorption, associated with dissipation due to spontaneous emission at rate $\gamma$, is negligible, so are the noise effects of vacuum fluctuations; Eq. (\ref{NLSE}) then holds in operator form without additional Langevin  quantum noise operators.

\subsection{Derivation of Eq. (\ref{NLSE})}
The formal derivation of Eq. (\ref{NLSE}) goes as follows (more details in Appendix A). The field envelope $\hat{\mathcal{E}}(z)=\sum_k \hat{a}_ke^{ikz}/\sqrt{L}$, with commutation relations $[\hat{a}_k,\hat{a}_{k'}]=\delta_{kk'}$ and hence $[\hat{\mathcal{E}}(z),\hat{\mathcal{E}}^{\dag}(z')]=\delta(z-z')$, is assumed to be spectrally narrow and is guided by a transverse mode of the waveguide (later taken to be the HE$_{11}$ mode of a fiber) with effective area $A$ at the atomic position $r_a$ and polarization vector $\mathbf{e}_0$.
The Hamiltonian in the interaction picture is (see also Appendix A1)
\begin{eqnarray}
H_{AF}= -\hbar n_{a}\int d z\left[i g \hat{\mathcal{E}}(z)e^{ik_0z}\hat{\sigma}_{eg}(z)+\mathrm{h.c.}\right],
\nonumber\\
H_{AC}= -\hbar n_a\int d z \left[i \Omega e^{-i\delta_c t} e^{ik_c z} \hat{\sigma}_{ed}(z)+\mathrm{h.c.}\right],
\nonumber\\
H_{DD}= \frac{1}{2}n_a^2\hbar \int dz\int dz' U(z-z') \hat{\sigma}_{dd}(z) \hat{\sigma}_{dd}(z'),
\label{HDD}
\end{eqnarray}
and $H_{F}=\sum_k\hbar c k\hat{a}^{\dag}_k \hat{a}_k$,
where $g=\sqrt{\omega_0/(2\epsilon_0\hbar A)}\mathbf{d}\cdot\mathbf{e}_0$, $\mathbf{d}$ being the dipole matrix element of the $|g\rangle\rightarrow|e\rangle$ transition, and $\hat{\sigma}_{ij}(z)=|i\rangle\langle j|$ for an atom at $z$, with $i,j$ representing the states $\{g,d,e\}$. By writing the Heisenberg equations for the atom and field operators and assuming a sufficiently weak probe field such that the atomic $|g\rangle\rightarrow |e\rangle$ transition is far from saturation, we obtain coupled equations for the $|g\rangle \leftrightarrow |d\rangle$ spin-wave and the field (Appendix A2),
\begin{eqnarray}
&&\bar{\sigma}_{gd}(z)=-\frac{g}{\Omega}\hat{\mathcal{E}}(z)-\frac{1}{\Omega}\left[\left(\partial_t+\gamma\right)\left(\frac{1}{\Omega^{\ast}}\right)\right.
\nonumber\\
&&\quad\quad\quad\quad \left.\times\left(\partial_t+i\delta_c+i\hat{S}(z)\right)\bar{\sigma}_{gd}(z)-\hat{F}\right],
\nonumber \\
&&\left(\partial_t+c\partial_z\right)\hat{\mathcal{E}}(z)=n_a \frac{g^{\ast}}{\Omega^{\ast}}\left[\partial_t+i\delta_c+i\hat{S}(z)\right]\bar{\sigma}_{gd}(z),
\nonumber\\
\label{EOM}
\end{eqnarray}
where $\bar{\sigma}_{gd}(z)=\hat{\sigma}_{gd}(z)e^{i(k_c-k_0)z}e^{-i\delta_c t}$. Here $\gamma$ and $\hat{F}$ are the spontaneous emission rate and corresponding Langevin noise operator, respectively due to the coupling of the $|g\rangle\rightarrow |e\rangle$ transition to the reservoir formed by photon modes other than those guided by the waveguide. The effect of interaction is a nonlinear detuning for the coupling field,
\begin{equation}
\hat{S}(z)=n_a\int_Ldz'U(z-z')\bar{\sigma}_{gd}^{\dag}(z')\bar{\sigma}_{gd}(z').
\label{S}
\end{equation}
Moving to the polariton picture of EIT \cite{FL}, we define the dark and bright polaritons, $\hat{\Psi}$ and $\hat{\Phi}$ respectively,
\begin{eqnarray}
\left(
  \begin{array}{c}
   \hat{\Psi}(z) \\
    \hat{\Phi}(z) \\
  \end{array}
\right)
=
\left(
  \begin{array}{cc}
    \cos\theta &-\sqrt{N}\sin\theta \\
    \sin\theta & \sqrt{N}\cos\theta \\
  \end{array}
\right)
\left(
  \begin{array}{c}
   \hat{\mathcal{E}}(z) \\
    \bar{\sigma}_{gd}(z)/\sqrt{L} \\
  \end{array}
\right),
\label{pol}
\end{eqnarray}
with $\tan^2\theta=n_a g^2/|\Omega|^2$, and transform Eqs. (\ref{EOM}) into equations of motion for the polaritons,
\begin{eqnarray}
&&\left(\partial_t+c\cos^2\theta \partial_z\right)\hat{\Psi}(z)=-\sin\theta\cos\theta c\partial_z\hat{\Phi}(z)
\nonumber \\
&&\quad\quad\quad\quad-i\sin\theta\left(\delta_c+\hat{S}(z)\right)\left[\sin\theta\hat{\Psi}(z)-\cos\theta\hat{\Phi}(z)\right],
\nonumber \\
&&\hat{\Phi}(z)=\frac{\cos\theta}{|\Omega|^2}(\partial_t+\gamma)\left(\partial_t+i\delta_c+i\hat{S}(z)\right)
\nonumber\\
&&\quad\quad\quad\quad\times\left[\sin\theta\hat{\Psi}(z)-\cos\theta\hat{\Phi}(z)\right]+\sqrt{n_a}\frac{\cos\theta}{\Omega}\hat{F}.
\label{EOMpol}
\end{eqnarray}

In the adiabatic regime, where the probe field is a CW and in the absence of detunings ($\delta_c,U=0$), the bright polariton $\hat{\Phi}$ vanishes \cite{FL}. Here, we take the first non-adiabatic correction (Appendix A3) by inserting the equation for $\hat{\Phi}$ into that of $\hat{\Psi}$, and assume all detunings to be smaller than the EIT transparency window $\delta_{tr}=\Omega^2/(\gamma\sqrt{OD})$, with $OD=(n_a/A) L \sigma_a$ and $\sigma_a$ the cross section of the $|g\rangle\rightarrow |e\rangle$ transition, finally arriving at Eq. (\ref{NLSE}) with $\alpha=\sin^2\theta$, $v=c\cos^2\theta$, $C=\sin^2\theta(2-3\sin^2\theta)/|\Omega|^2$.

\section{Collective excitations and propagation in a CW background}
\subsection{Excitation spectrum of polariton waves}
With Eq. (\ref{NLSE}) in hand, we now turn to the analysis of the polariotn wave propagation it predicts.
Specifically, let us assume a CW polariton, and find the dispersion relation of wave excitations around this CW background, in analogy to the Bogoliubov spectrum of excitations in a Bose-Einstein condensate (BEC) \cite{PS,SP}.
The CW solution of (\ref{NLSE}) is $\psi(t)=\psi_0e^{-i(\alpha\delta_c+ n_p U_0) t}$, with $\psi_0=|\psi_0|e^{i\phi}$, $n_p=\alpha^2 |\psi_0|^2$ being the effective photon density per unit length and $U_0$ the $k=0$ component of the spatial Fourier transform of the potential $U_k=\int_{-\infty}^{\infty}dz U(z)e^{-ikz}$.
Here we have neglected edge effects by assuming $l<z<L-l$, $l$ being the range of the potential $U(z)$. The dispersion relation of small fluctuations $\varphi(z,t)$ around the large average CW field $\langle\Psi\rangle=\psi(t)$ are found as usual upon inserting $\Psi=\psi(t)+\varphi(z,t)$ into Eq. (\ref{NLSE}) and linearizing it by keeping the fluctuations $\varphi$ to linear order.
Then, introducing the ansatz \cite{PS},
\begin{eqnarray}
\varphi(z,t)&=&e^{i[\phi-(\alpha\delta_c+n_p U_0)t]}
\nonumber\\
&\times&\left[u_ke^{ikz}e^{-i(\omega_k+kv) t} - v_k^{\ast} e^{-ikz}e^{i(\omega_k+kv) t}\right],
\label{an}
\end{eqnarray}
into the linearized equation for $\varphi$, $u_k$ and $v_k$ being c-number (Bogoliubov) coefficients, and using standard procedures, we find the modified Bogoliubov spectrum (Appendix B1)
\begin{equation}
\omega_k=\sqrt{\omega_k^0(\omega_k^0+2n_pU_k)}, \quad \omega_k^0=(n_pU_0/\alpha+\delta_c)Cv^2k^2.
\label{bog}
\end{equation}
This means that a polariton wave distortion (about the CW solution) with a wavenumber $k$ relative to the carrier wavenumber (inside the EIT medium), oscillates at a frequency (relative to the carrier frequency $\omega_0$)
\begin{equation}
\omega(k)=\alpha\delta_c+n_p U_0 \pm (v k + \omega_k),
\label{DR}
\end{equation}
where the $\pm$ sign is for positive/negative $k$, respectively. The dispersion relation (\ref{DR}) is composed of the detuning $\alpha\delta_c$ due to the coupling field, the self-phase modulation of the CW component $n_p U_0$ (analogous to the chemical potential in a BEC \cite{PS}), the linear dispersion $vk$, and the modified Bogoliubov excitation spectrum $\omega_k$. The spectrum $\omega_k$ is determined by the interplay between the interaction Fourier transform $U_k$ and the "kinetic-energy" quadratic dispersion $\omega_k^0$, which is affected by both the detuning $\delta_c$ and by the $k=0$ component of $U_k$. This interplay is further discussed below for $U(z)$ resulting from laser-induced interactions near a waveguide grating.

\subsection{Generation of two-mode correlations}
The parametric process described by the foregoing modified Bogoliubov theory also entails the dynamic generation of two-mode squeezing, i.e. pairs of entangled polaritons with wavenumbers $\pm k$. The analysis is similar to that of propagation in fibers with local Kerr nonlinearity \cite{YP,DF}. Upon inserting the expansion of small quantum fluctuations in the longitudinal wavenumber modes $k$, $\hat{\varphi}(z)=\sum_k \hat{a}_k e^{ikz}/\sqrt{L}$, into the linearized equation for $\varphi(z,t)$, we obtain coupled first-order differential equations (in time) for $\hat{a}_k(t)$ and $\hat{a}^{\dag}_{-k}(t)$, whose solution is a dynamic Bogoliubov transformation (Appendix B2),
\begin{eqnarray}
\hat{a}_k(t)&=&e^{-i(n_pU_0+\alpha\delta_c+kv)t}\left[\mu_k(t)\hat{a}_k(0)+e^{i2\phi}\nu_k(t)\hat{a}^{\dag}_{-k}(0)\right],
\nonumber\\
\mu_k(t)&=&\cos(\omega_k t)-i\frac{n_pU_k+\omega_k^0}{\omega_k}\sin(\omega_k t),
\nonumber\\
\nu_k(t)&=&-i\frac{n_pU_k}{\omega_k}\sin(\omega_k t).
\label{bog2}
\end{eqnarray}
The number of entangled pairs generated after propagation time $t$ at wavenumbers $\pm k$ can be quantified by the so-called squeezing spectrum, whose optimum is given by $G_k=(|\mu_k|-|\nu_k|)^2$ \cite{YP} ($G_k<1$ signifies entanglement), or by the normalized second-order (intensity) correlation function $g^{(2)}(z,z',t)=[\langle\hat{\Psi}^{\dag}(z)\hat{\Psi}^{\dag}(z')\hat{\Psi}(z')\hat{\Psi}(z)\rangle]/[\langle\hat{\Psi}^{\dag}(z)\hat{\Psi}(z)\rangle\langle\hat{\Psi}^{\dag}(z')\hat{\Psi}(z')\rangle]$ (all fields measured at the  waveguide's output after propagation time $t=L/v$), where the averaging is performed with respect to the initial probability distribution, e.g. the quantum state, of the polariton (probe) field. For initial zero-mean fluctuations (around the CW solution) with average polariton occupation at mode $k$, $\langle \hat{a}_k^{\dag}(0)\hat{a}_{k'}(0)\rangle=N_k\delta_{kk'}$, and vanishing anomalous correlations $\langle \hat{a}_k(0)\hat{a}_{k'}(0)\rangle=0$, we find (Appendix B3)
\begin{eqnarray}
&&g^{(2)}(z,z')\approx 1+\frac{2\alpha^2}{\pi n_p}\int_{0}^{\infty}dk \left[|\mu_k|^2N_k+ |\nu_k|^2(N_k+1)\right.
\nonumber\\
&&\left.+(2N_k+1)|\mu_k||\nu_k|\cos\phi_k\right]\cos[k(z-z')],
\nonumber\\
\label{g2}
\end{eqnarray}
where $\phi_k=\mathrm{arg}(\mu_k\nu_k)$ and we used $(1/L)\sum_k\rightarrow\int dk/(2\pi)$. As we shall see below, these correlations may reveal the ordering of a nonlocal system caused by pair-generation at preferred $k$-values.

\section{Highly nonlocal laser-induced interaction via waveguide grating}
Our analysis up to this point was kept general, without specifying the interaction potential $U(z)$. Let us now turn to a particulary interesting case of an extremely long-range interaction, where novel nonlinear optical effects can be illustrated.
\subsection{Shaping the interaction potential}
The illumination of atoms by an off-resonant laser virtually excites the atoms and allows them to resonantly interact via virtual photons. The spatial dependence of the resulting interaction potential $U(z)$ then follows the spatial structure of the mediating photon modes. This is the essence of the laser-induced interaction potential we wish to employ \cite{LIDDI}. Specifically, consider another laser with Rabi frequency $\Omega_L$ which is detuned by $|\delta_L|\gg\Omega_L$ from the transition $|d\rangle\rightarrow |s\rangle$, $|s\rangle$ being a fourth atomic level (Fig. 1c), and assume that this transition is distinct and separated from the transitions used for EIT (Fig. 1b) either spectrally or by polarization. Then, the extended waveguide modes can mediate long-range interactions between the trapped atoms (Fig. 1a) (Appendix C). As a specific example, we consider a nano-waveguide that incorporates a grating, i.e. periodic perturbation of the refractive index with period $\Lambda\equiv\pi/k_B$, so that the photon modes exhibit a bandgap (see e.g. Refs. \cite{KIM2,HAK}). Then, for a laser frequency $\omega_L$ inside the gap and close to its edge (the probe field's carrier frequency $\omega_0$ being outside the gap), the laser-induced interaction potential becomes \cite{LIDDIna} (Appendix C)
\begin{equation}
U(z)=-U_L\frac{1}{2}\cos(k^z_L z)\cos(k_B z)e^{-|z|/l},
\label{UL}
\end{equation}
where $l$ can extend over hundreds of wavelengths \cite{LIDDIna,CHA}. Here $k^z_L=k_L \cos\theta_L$ with $k_L$ the laser wavenumber and $\theta_L$ its orientation with respect to the waveguide axis $z$ (Fig. 1c), and $U_L$ depends on the laser parameters, atomic transition and effective area at $r_a$ (see Appendix D). The resulting spatial Fourier transform $U_k$ then consists of four Lorentzian peaks of width $\sim 1/l$, centered around the spatial beating frequencies $\pm(k^z_L-k_B)$ and $\pm(k^z_L+k_B)$.
We note that the laser $\Omega_L$ and the grating are unrelated to the linear propagation of the probe field and their sole role is to induce the long-range dipolar interaction Eq. (\ref{UL}) between atoms, which is in turn translated via EIT to interaction between polariotns as in Eq. (\ref{NLSE}) (see also Discussion, Sec. VII B).

\subsection{Roton and Anti-Roton spectra}
Let us focus on the peak of $U_k$ around $k_R\equiv k^z_L-k_B$ and its effect on the dispersion relation (spectrum), Eq. (\ref{bog}). We first consider the case  of anomalous dispersion, where the signs of $\omega_k^0$ and $U_k$ are opposite. In analogy to BEC, this describes the case of an attractive potential $U_k$ that competes with the "kinetic energy"  $\omega_k^0$. Then, for $k$-values satisfying $|\omega_k^0|>2n_p|U_k|$, $\omega_k$ is real and exhibits a dip around $k_R$, in contrast to the case of a local potential for which $U_k$ is independent of $k$ (standard Bogoliubov spectrum) and this feature is absent. This is seen in Fig. 3a, where both the analytical results of Eq. (\ref{bog}) and numerical simulations of the nonlinear equation (\ref{NLSE}) (Appendix \ref{ANUM}), are plotted and shown to agree very well. The narrow-band "dip" of this $\omega_k$ spectrum is in analogy with the roton minimum in He II \cite{PN}. It reflects the fact that wave distortions about the CW field with spatial frequencies around $k_R$ cost less energy and are hence favorable. This feature implies that the intensity of the polariton field in its ground state would tend to self-order with typical wavenumber $k_R$ \cite{PN}.

Turning to the case of normal dispersion, where the signs of $\omega_k^0$ and $U_k$ are identical, $\omega_k$ exhibits an "anti-roton" peak around $k_R$ (Fig. 3b). This means that distortions of spatial frequencies around $k_R$ are costly, so that the system prefers to avoid these spatial variations. This behavior again manifests the tendency of the system to order, since it indicates the spatial distortions that the system is unlikely to be found in, should it be in its ground state.
\begin{figure}
\begin{center}
\includegraphics[scale=0.30]{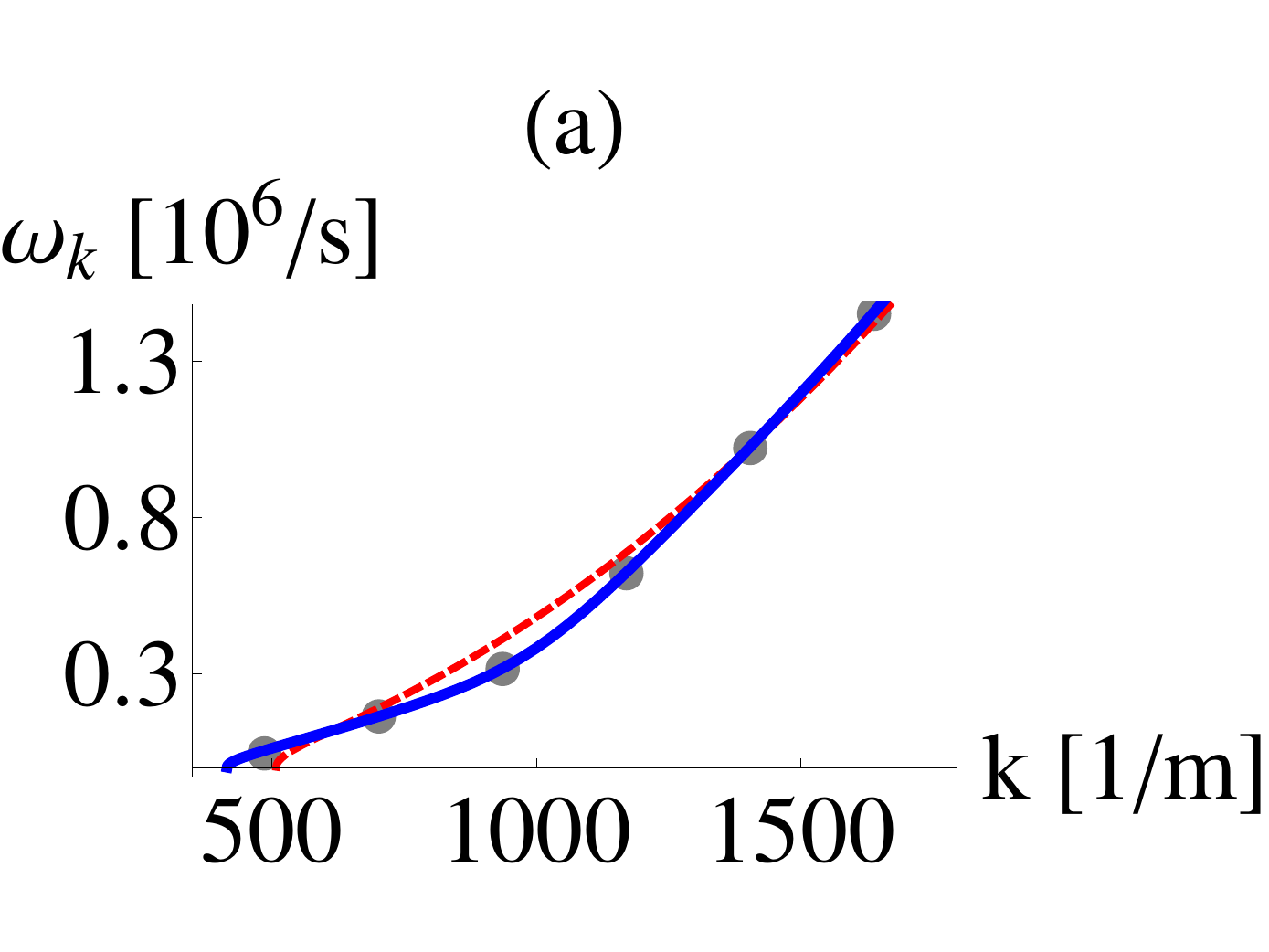}
\includegraphics[scale=0.30]{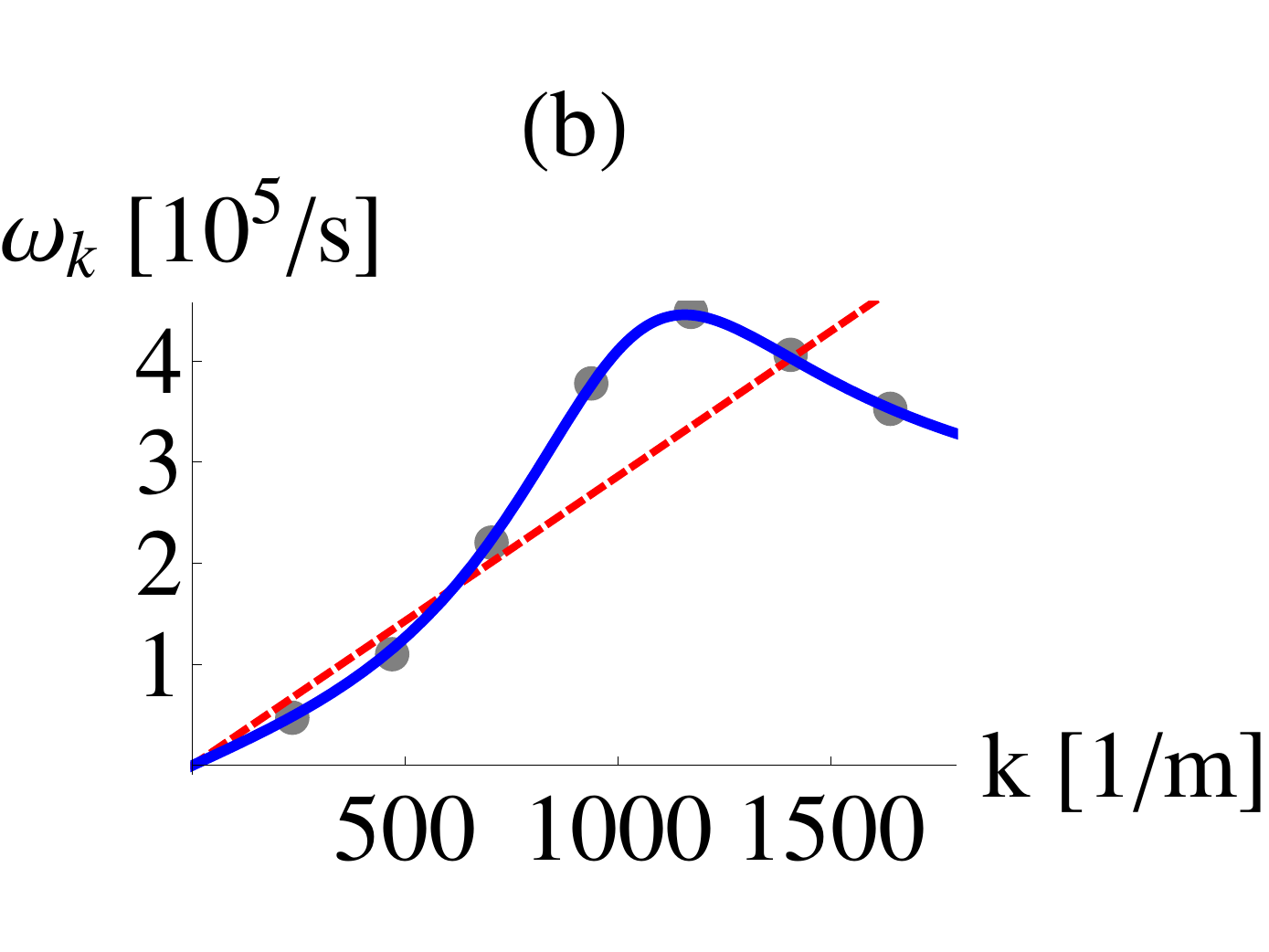}
\includegraphics[scale=0.36]{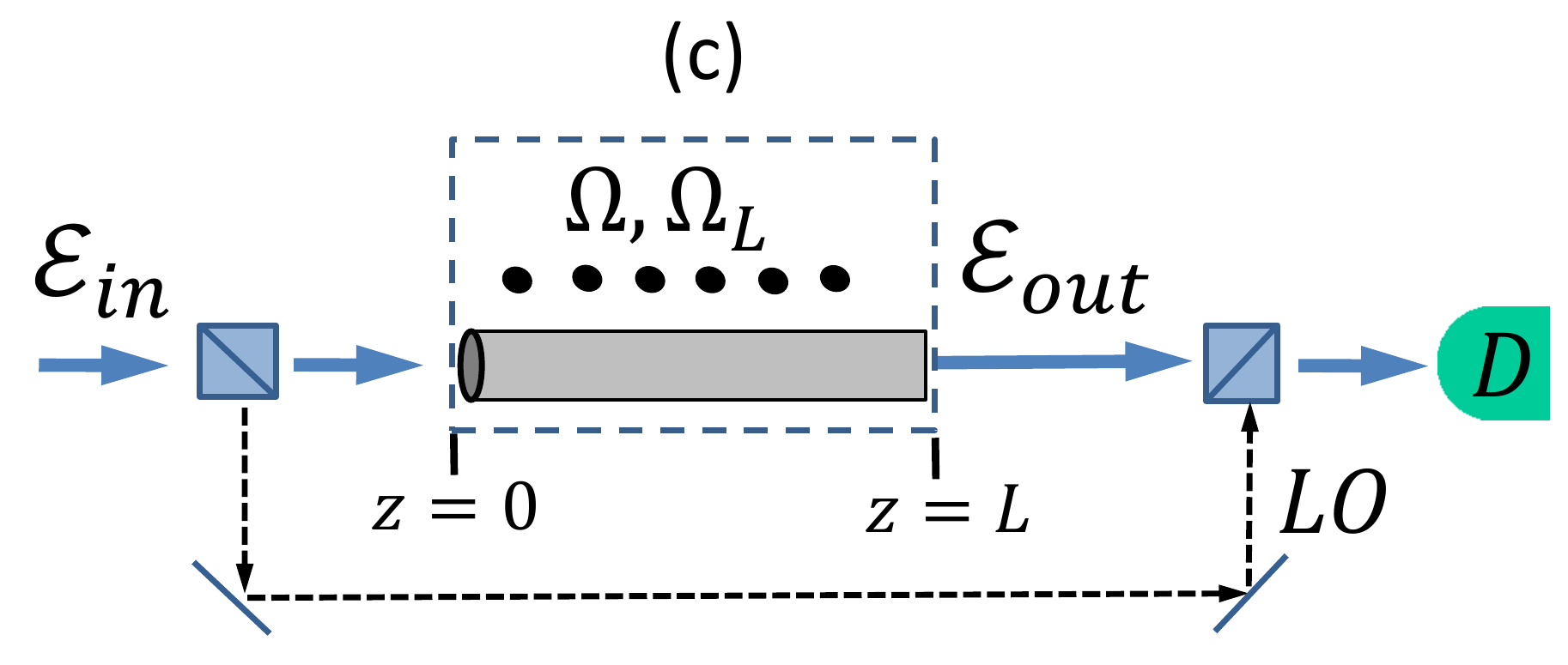}
\caption{\small{
Dispersion relations for EIT polaritons with waveguide-grating mediated atomic interactions ($k$ values presented in all plots are within the EIT transparency window; values of physical parameters used here are given in Appendix \ref{AQUA}). (a) Roton-like excitation spectrum (dispersion relation) $\omega_k$ for the potential of Eq. (\ref{UL}) in the anomalous dispersion case (opposite signs of $\omega_k^0$ and $U_k$). The analytical results from Eq. (\ref{bog}) (blue solid line) agree well with those of direct numerical simulations of Eq. (\ref{NLSE}) (gray dots). Compared with the spectrum of a local interaction ($U_k$ independent of $k$, dashed red line), the roton-like spectrum exhibits a dip in a narrow band of $k$-values around $k_R=k_L^z-k_B$. (b) Anti-roton peak of the spectrum $\omega_k$ in a narrow band around $k_R$ in the normal dispersion case (identical signs of $\omega_k^0$ and $U_k$). (c) Possible homodyne detection scheme: the input probe field consists of a CW field + perturbation at wavenumber $k$ and frequency $\omega^{(0)}(k)$. The field is split before entering the EIT medium ($z=0$), so that a local oscillator of $\omega^{(0)}(k)$ is formed (lower arm) by e.g. filtering out the CW component. Then, mixing the output signal ($z=L$) with the local oscillator reveals their phase difference, from which $\omega_k$ can be inferred (see text).
}}
\end{center}
\end{figure}

In order to measure the roton and anti-roton spectra, we first recall the meaning of the dispersion relation $\omega(k)$ from Eq. (\ref{DR}): without an interaction, the frequency associated with a wave envelope at wavenumber $k$ traveling inside the EIT medium is given by $\omega^{(0)}(k)=\alpha\delta_c+vk+\delta_c C v^2 k^2=\omega(k)-n_p U_0-(\omega_k-\delta_c C v^2 k^2)$, so that $\omega_k$ (together with $n_pU_0$) expresses a frequency, or phase velocity, shift due to the nonlinearity. Namely, the wavenumber $k$ describes a spatial eigenmode of propagation, both with and without interaction, with an eigenfrequency $\omega(k)$ and $\omega^{(0)}(k)$, respectively. Now, suppose we let a weak quasi-CW pulse of length $L_p<L$ and frequency $\omega_p=\omega^{(0)}(k)$ enter the medium (on top of the strong CW), when the laser and hence the interaction $U_k$ are turned off. Since upon entering the medium the field does not change its frequency $\omega_p$, we deduce from the dispersion relation in the absence of interaction, $\omega^{(0)}(k)$ that the field inside the medium exhibits a perturbation $\varphi(z)$ at spatial frequency $k$ on top of the strong CW. Subsequently, when the entire pulse is in the medium, we immediately (non-adiabatically) turn on the laser $\Omega_L$, and hence the interaction $U_k$, so that the temporal frequency of the perturbation $\varphi(z)$ at wavenumber $k$ becomes $\omega(k)=\omega^{(0)}(k)+n_pU_0+(\omega_k-\delta_c C v^2 k^2)$. The frequency shift $n_pU_0+(\omega_k-\delta_c C v^2 k^2)$ can be thought of as an extra energy acquired by the mode $k$ due to the interaction energy. Therefore, the spectrum $\omega_k$ can be inferred from the frequency shift, measurable by homodyne detection of the pulse $\varphi(z)$ that exits the EIT medium (Fig. 3c). A similar procedure was proposed for measuring the tachyon-like spectrum of polaritons in inverted media \cite{TACH}.

\subsection{Dynamical instability: pair generation}
In the anomalous dispersion case, consider now a sufficiently strong interaction such that for a narrow band of $k$-values around $k_R$, where $U_k$ is peaked, the condition $2n_p|U_k|>|\omega_k^0|$ is satisfied and $\omega_k$ becomes imaginary. Then, field perturbations around $k_R$ become exponentially unstable (Fig. 4a), resulting in parametric amplification and generation of entangled photon pairs in this narrow band of unstable $k$-values. The strength of the amplified perturbation and generated field is characterized by the magnitude of the coefficients $\mu_k(t)$ and $\nu_k(t)$ from Eq. (\ref{bog2}), which grow exponentially with propagation time $t$ and are largest for the narrow peak around $k_R$. The resulting squeezing spectrum $G_k$ at the output $t=L/v$ (Fig. 4a) may be measured by homodyne detection \cite{YP}.

\begin{figure}
\begin{center}
\includegraphics[scale=0.30]{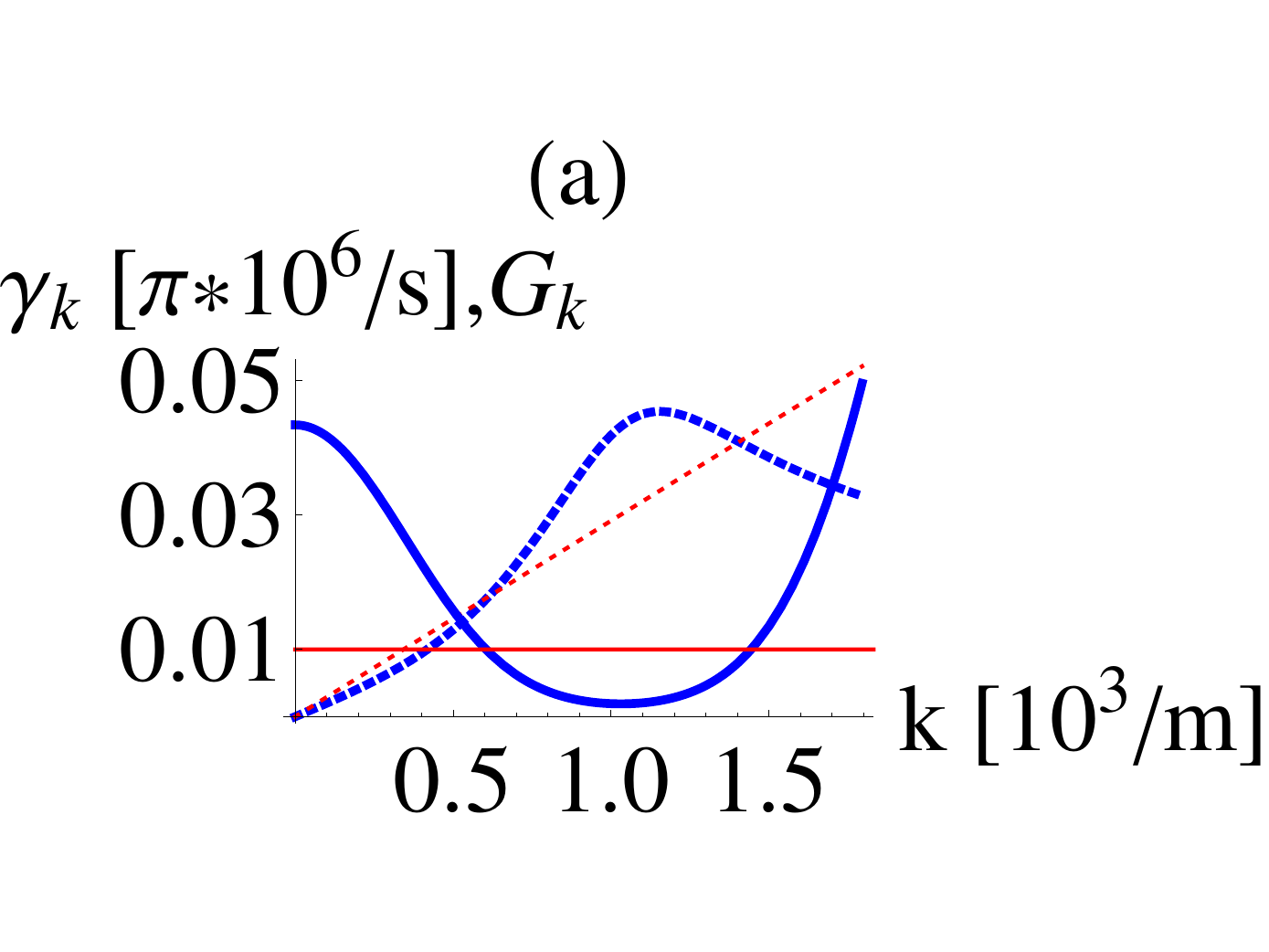}
\includegraphics[scale=0.30]{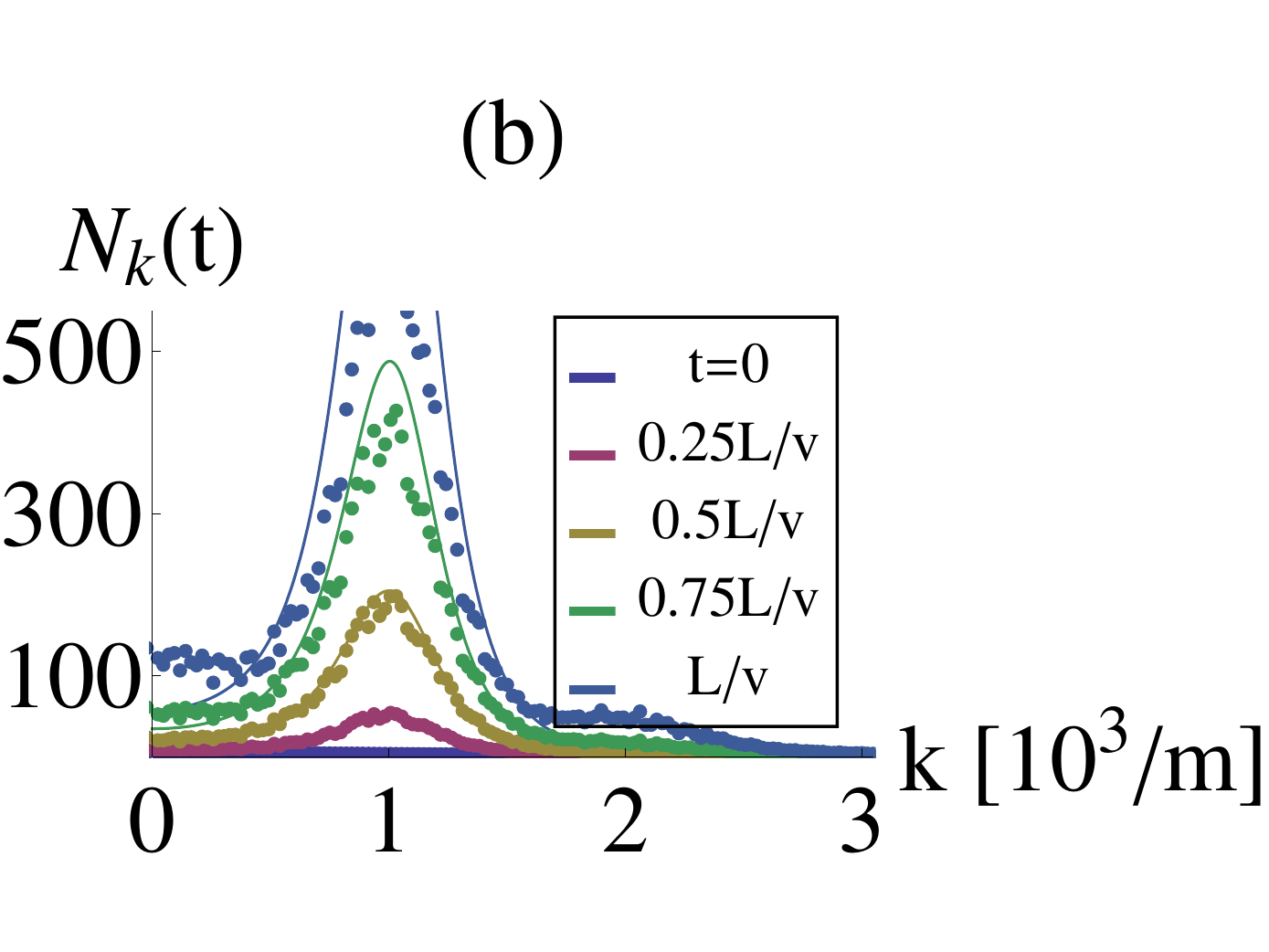}
\includegraphics[scale=0.33]{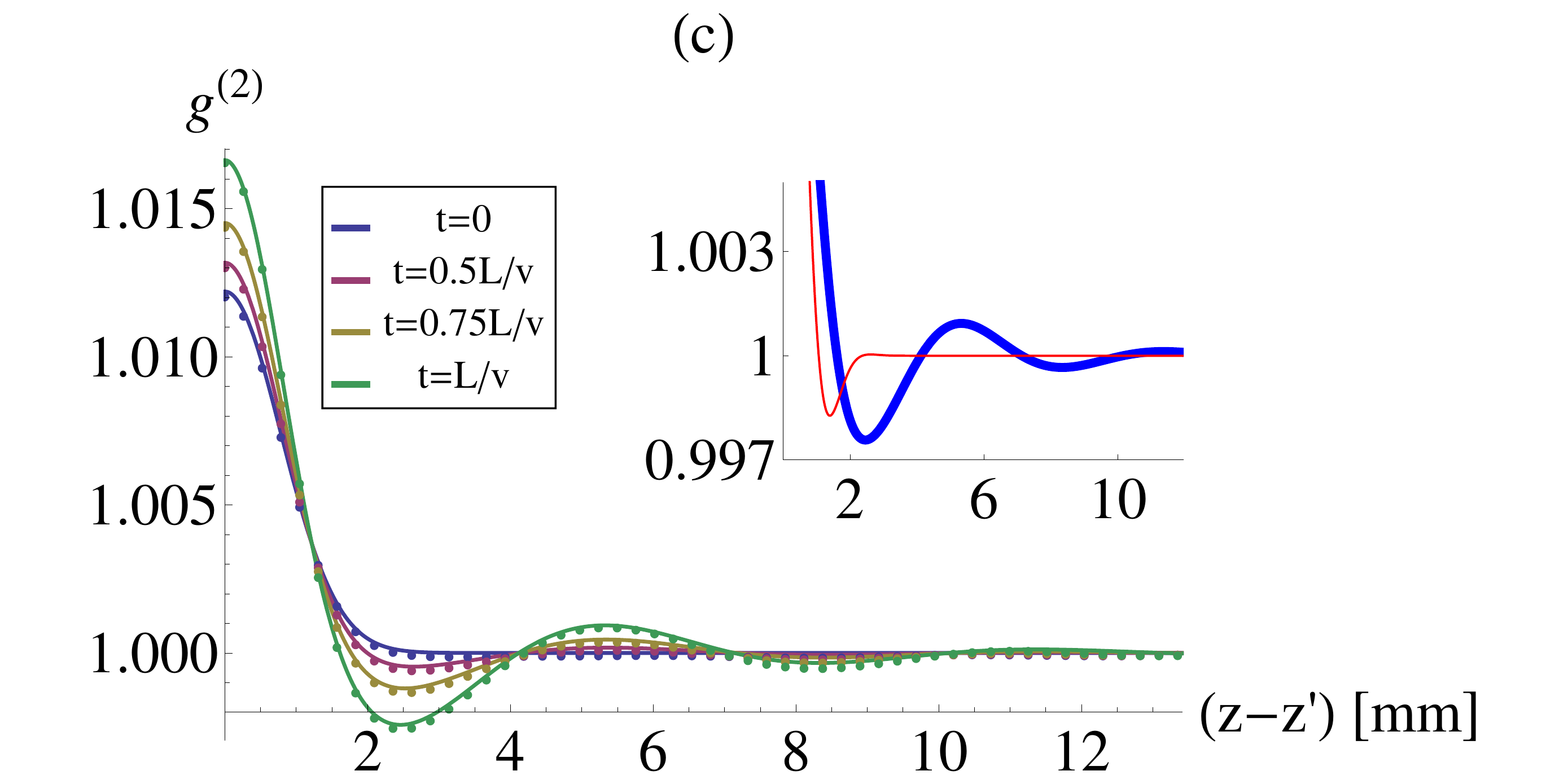}
\caption{\small{Instability and self-ordering (values of physical parameters are given in Appendix \ref{AQUA}).
(a) $\gamma_k=\mathrm{Im}\omega_k$ is the exponential growth rate of unstable perturbations at spatial frequency $k$ (anomalous dispersion case). It exhibits a narrow peak around $k_R$ (blue dotted line), compared to the broadband instability of the local-interaction case (red dotted line). The instability is accompanied by generation of quantum entanglement, characterized by a narrow-band squeezing spectrum $G_k$ (blue solid line; $G_k<1$ quantifies entanglement between $\pm k$ photon modes), in contrast to the broadband spectrum for a local interaction (horizontal red solid line).
(b) Dynamics of spatial spectrum $N_k(t)$ of the field intensity for different propagation times $t$ inside the medium ($t=L/v$ at the output). The emergence of a large peak around $k_R$ (and $-k_R$) out of the initial Gaussian perturbation is clearly seen. Results of both the linearized theory (solid lines) and numerical simulations of Eq. (\ref{NLSE}) (dots) are shown: slight differences at later $t$-values are attributed to nonlinear corrections.
(c) Self-ordering of the field intensity: considering the narrow peaks of $N_k(t)$ around $\pm k_R$, fluctuations at these $k$-values become dominant, resulting in ordered intensity correlations $g^{(2)}$ which grow with propagation time $t$ ($t=L/v$ at the output). Excellent agreement between the theory, Eq. (\ref{g2}), (solid lines) and numerical simulations of Eq. (\ref{NLSE}) (dots) is observed. The correlations oscillate with a period $2\pi/k_R\sim6.1$ mm and a range of a few $l\approx 2.7$ mm (where $L=2.68$ cm and $v=4340$ ms$^{-1}$), determined by the long-range interaction $U(z)$. This is in contrast to the local-interaction case, where the correlations vanish ($g^{(2)}=1$) after a short distance $\pi/q_{tr}\approx1.75$ mm, which is determined by the bandwidth $q_{tr}$ of the initial fluctuations [see inset: local (red thin line) and nonlocal (blue thick line) cases for the output field at $t=L/v$].
}}
\end{center}
\end{figure}
\subsection{Dynamical instability: emergence of self-order}
An interesting implication of the extremely nonlocal potential of Eq. (\ref{UL}) is the \emph{dynamic formation of order} in the system. Consider that at $t=0$ there exist fluctuations around the CW, with a spatial spectrum $N_k=N_0 e^{-(k/q_{tr})^2}$, i.e. "$\delta$-correlated" noise limited by the EIT transparency window of width $\delta_{tr}=v q_{tr}$.
Then, fluctuations at $k$-values around the peak $k_R$ will be parametrically amplified as they propagate through the medium, as verified in Fig. 4b, where the spatial spectrum of the polariton field, $N_k(t)=\langle\hat{a}_k^{\dag}(t)\hat{a}_k(t)\rangle$, at different propagation times $t$ ($t=L/v$ describing the output field) is calculated both analytically and via direct (classical) numerical simulations of Eq. (\ref{NLSE}). This suggests that the system becomes spatially ordered with a period $\sim2\pi/k_R$, which may be revealed by measuring the correlation function $g^{(2)}$ between the intensities of the output field that arrive at a detector at the waveguide's end ($z=L$) and time difference $(z-z')/v$ (scheme from Fig. 3c without the lower local-oscillator arm). We obtain the corresponding $g^{(2)}$ by numerically integrating over $k$ in the classical limit of Eq. (\ref{g2}), where the vacuum-fluctuations contributions are neglected. This calculation is compared to $g^{(2)}$ measured via direct numerical simulations of Eq. (\ref{NLSE}), yielding excellent agreement. Fig. 4c reveals the emergence of order by presenting the intensity-correlations $g^{(2)}$ at different propagation times $t$ through the medium ($t=L/v$ at the output): The resulting $g^{(2)}$ exhibits oscillations with a period $z-z'\sim 2\pi/(k_R)$, which persist over a few $l$. Considering that the interaction range $l$ can reach hundreds or even thousands of optical wavelengths ($l\approx3027\lambda_L$ and $L\sim10l\sim0.026$ m in our example, see Appendix \ref{AQUA}), the light intensity clearly  becomes ordered due to the long-range interaction $U(z)$, as can also be seen by the comparison to the local-interaction case (inset of Fig. 4c).

\section{Scattering and imperfections}
So far we have considered a purely coherent evolution of the polariton field. In Appendix \ref{ASC} we address three main sources of scattering and loss of field excitations; we estimate the decoherence rate each of them imposes on the polariton field and its possible effect on the observability of the self-ordering effects discussed above.

First, since the interaction $U(z)$ from Fig. 1b is induced by the illumination of the atoms by an off-resonant laser, it is accompanied by an incoherent process of scattering of laser photons $\Omega_L$ from the $|d\rangle\rightarrow |s\rangle$ transition to non-guided modes at rate $R_{fs}$ \cite{LIDDI}. This process limits the coherence time of the $\bar{\sigma}_{gd}$ spin wave and hence that of the polariton to be below $\sim R_{fs}^{-1}$, which in the example of Figs. 3 and 4 is nevertheless much longer than the experiment time $L/v$ (Appendix \ref{AQUA}).

Second, consider the losses due to propagation in the EIT medium with a non-vanishing detuning of the coupling field, $\Delta_c=\delta_c+\delta_{NL}$, where here $\delta_{NL}= n_pU_0/\alpha$ is the mean-field value of $\hat{\delta}_{NL}$. The transmission through the EIT medium of length $L$ is given by $e^{-(1/2)k_0 \chi'' L}$, leading to a decoherence rate  $R_{EIT}\sim(1/2)k_0 v \chi''$ for the polariton, where $\chi''=\mathrm{Im}\chi(\Delta_p=0,\Delta_c)$ is the imaginary part of the EIT susceptibility \cite{FIM} evaluated for simplicity at the center of the probe pulse ($\Delta_p=0$). In Appendix \ref{ASC}2 we show that the effect of the EIT losses on the observability of $\omega_k$ can be significantly decreased by reducing both $n_p$ and $\delta_c$, resulting in a loss rate smaller than the main features of the $\omega_k$ (roton and anti-roton cases) and $\gamma_k=\mathrm{Im}\omega_k$ (instability case) spectra.

Finally, we consider material imperfections in the waveguide grating structure (e.g. defects and surface roughness) which give rise to scattering of photons off the guided modes. This leads to a decay rate (width) $\kappa$ for each of the longitudinal (guided) photon-modes $k$ (see Appendix \ref{ASC}3 and \cite{CHA,LUK}). Then, since atoms at level $|d\rangle$ are coupled via $\Omega_L$ to these lossy waveguide-grating modes, the $\bar{\sigma}_{gd}$ spin wave and hence the polariton are decohered at a rate $R^{(1)}_{im}\sim(\kappa/\delta_u)U_L$, $\delta_u$ being the detuning of the laser $\Omega_L$ from the upper bandedge of the grating.
However, the effect of $R^{(1)}_{im}$ on the spectra $\omega_k$ and the observables mentioned above can in principle be made arbitrary small, by noting that $\omega_k$ depends on $n_p U_L$ whereas $R^{(1)}_{im}$ depends solely on $U_L$, reflecting the fact that $R^{(1)}_{im}$ is a single-polariton loss mechanism whereas $\omega_k$ is a cooperative effect. Then, decreasing $|U_L|$ while keeping $n_p U_L$ constant (by increasing the CW power $n_p$) reduces $R^{(1)}_{im}$ but keeps $\omega_k$ unchanged.

Nevertheless, a cooperative decoherence can in fact result from $\kappa$ and $U(z)$; namely, a pair of atoms, at a distance $z$ apart, can jointly scatter photons at a rate $\sim(\kappa/\delta_u)U(z)$, since the virtual photons that mediate their interaction $U$ are now lossy and enable excitation decay to non-guided modes. Then, for a single polariton at $z$, the scattering induced by the entire atomic medium becomes $R^{(2)}_{im}\sim n_p\int dz'(\kappa/\delta_u)U(z')\sim (\kappa/\delta_u)n_p U_0$. In Appendix \ref{ASC}3 we show that the limitation imposed by this dephasing channel on the observability of $\omega_k,\gamma_k$ can be reduced by increasing the EIT coupling laser $\Omega$.

In principle, all loss terms discussed above should be accompanied by corresponding quantum noise terms, which are not considered here, restricting this discussion to the classical-field case (for which all of the above results apply, excluding the squeezing spectrum $G_k$).

\section{Discussion}
This study predicts a new and hitherto unexplored regime of nonlinear optics; namely, that of highly nonlocal interactions between photons in one dimension (1d). These nonlocal optical nonlinearities arise for light propagation inside driven atomic media in the vicinity of a waveguide, and affect both the frequency and quadratic dispersion of the light field.
We have derived the nonlinear equation that governs light propagation in this regime, and have analyzed it around its CW solution, finding a narrow roton-like dispersion relation (Figs. 3a,b) and squeezing (entanglement) spectrum (Fig. 4a), and the emergence of order in the field-intensity (Figs. 4b,c), all of which reflect the tendency of the system to self-organize, which in turn results from the long-range interactions between photons. In the following we wish to discuss some important aspects of this work.

\subsection{Structure and generality of Eq. (\ref{NLSE})}
It is important to note the \emph{nonlocal and nonlinear dispersion} term appearing in Eq. (\ref{NLSE}), absent in previous works on nonlocal nonlinear optics
\cite{KRO,HD2,LC,AD1,AD2} and EIT-based nonlinear optics \cite{FRD,EFI,GOR,HE1,PF,POH,HE2,GOR2}. We identify the conditions under which this term becomes significant and may lead to new phenomena, namely, in the Autler Townes regime of EIT and for sufficiently large nonlinear detuning.

Another interesting point concerns the generality of our work in relation to \emph{different confining geometries}. The derivation of Eq. (\ref{NLSE}) does not depend on the specific form of the potential $U(z)$, which naturally opens the way to the exploration of nonlocal nonlinear optics due to laser-induced dipolar potentials mediated by confined photon modes of geometries other than the waveguide grating considered here. For example, the laser-induced potential in a cavity along $z$ is in general not translational-invariant and depends on the $z$-positions of both interacting atoms (rather than just their difference), a case which is qualitatively different from the one analyzed here.

\subsection{Origin and length-scale of self ordering}
We stress that the length-scale associated with order, $k_R=k_L\cos\theta_L-k_B$ originates from the interaction potential of the light field with itself, Eq. (\ref{UL}), hence ordering spontaneously occurs in this optical system much like in other condensed-mater systems and crystals. This is in contrast to e.g. order of atoms in a deep-potential optical lattice, where the atoms are situated at lattice sites determined by the potential imposed by an external laser rather than by their mutual interactions. In our case, the role of the grating is not to trap the propagating polariton but merely to create dispersive and long-range dipolar interaction between atoms $U(z)$ (induced by an external laser $\Omega_L$ that is unrelated to the light-component of the polariotn), which underlies the nonlinearity in the polariton propagation.

Moreover, it is important to note that the length-scale $k_R$ is a signature of the specific spatial shape of the potential $U(z)$ [sinusoidal in the case of Eq. (\ref{UL})], as also revealed by the excitation and instability spectra of Figs. 3 and 4(a). This is in contrast to e.g. the polariton crystallization process in a \emph{short-range} potential described in Ref. \cite{OTT} using Luttinger-liquid theory, where the specific shape of the potential is irrelevant. In this respect, our results are more related to the modulational instability discussed in Ref. \cite{POH} for Rydberg-atom EIT with a power-law potential, though there the considered system is three-dimensional and the quadratic dispersion is linear and emerges simply due to diffraction. Other recent works where photon spatial correlations follow those of the interaction potential include Refs. \cite{OFER2} and \cite{CHA3,CHA4}, which however treat a probe field with only a few photons (typically two), in contrast with the largely collective behavior discussed in the present work.

\subsection{Prospects}
To conclude, this work opens the way to experimental and theoretical investigations of new nonlinear wave phenomena, especially by venturing beyond the linearized regime and exploring the role of the nonlocal and nonlinear dispersion term $\hat{\delta}_{NL}Cv^2\partial_z^2$.
Specific directions of further research may include: \emph{Nonlocal nonlinear optics in 1d}, concerning the study of solitons, where the lensing effect created by the nonlinear refractive index change is now highly nonlocal, following $U(z)$; \emph{Thermalization in 1d}, which has been considered both experimentally \cite{SCH} and theoretically \cite{MAZ} for isolated BEC in 1d, may become qualitatively different here due to the nonlocal character of the designed interactions $U(z)$ and the nonlinear dispersion ("mass"); \emph{Effects of non-additivity} of systems with long-range interactions \cite{RUF} may be studied here for quantum/classical optical fields.

\acknowledgements
We acknowledge discussions with Ofer Firstenberg, Arno Rauschenbeutel, Philipp Schneewei{\ss}, Darrick Chang, Nir Davidson, Tommaso Caneva and Boris Malomed and the support of ISF, BSF and FWF [Project numbers P25329-N27, SFB F41 (VICOM) and I830-N13 (LODIQUAS)].

\appendix

\section{EIT with nonlocal interactions}
Here we elaborate on the derivation of Eq. (\ref{NLSE}). Sec. 1 and 2 below review rather standard results, whereas the adiabatic expansion leading to nonlocal and nonlinear dispersion is discussed in Sec. 3 and 4.
\subsection{Effective 1d description}
The system from Fig. 1 is modeled by the Hamiltonian
\begin{eqnarray}
H_A&=&\int d\mathbf{r} \rho(\mathbf{r})\left[\hbar\omega_{eg}\hat{\sigma}_{ee}(\mathbf{r})+\hbar(\omega_{eg}-\omega_{ed})\hat{\sigma}_{dd}(\mathbf{r})\right],
\nonumber\\
H_F&=&\sum_k\hbar(\omega_0+ck)\hat{a}_k^{\dag}\hat{a}_k,
\nonumber\\
H_{AF}&=& -\hbar \int d\mathbf{r} \rho(\mathbf{r}) \sum_k \left[i g_k(\mathbf{r}) \hat{a}_k e^{ik_0z}\hat{\sigma}_{eg}(\mathbf{r})+\mathrm{h.c.}\right],
\nonumber\\
H_{AC}&=& -\hbar \int d\mathbf{r}  \rho(\mathbf{r}) \left[i \Omega e^{-i\omega_c t} e^{ik_c z} \hat{\sigma}_{ed}(z)+\mathrm{h.c.}\right],
\nonumber\\
H_{DD}&=&\frac{1}{2}\int d\mathbf{r}\rho(\mathbf{r})\int d\mathbf{r}'\rho(\mathbf{r}') U(\mathbf{r}-\mathbf{r}') \hat{\sigma}_{dd}(\mathbf{r}) \hat{\sigma}_{dd}(\mathbf{r}'),
\nonumber\\
\label{H}
\end{eqnarray}
where $\hat{\sigma}_{ij}(\mathbf{r})=|i\rangle\langle j|$ are the continuum (coarse-grained) atomic operators around $\mathbf{r}$ \cite{FL}, and $\rho(\mathbf{r})$ is the atomic density. Here $\{k\}$ denote longitudinal modes guided by the waveguide with mode function $\mathbf{f}_k(x,y)e^{ikz}/\sqrt{L}$ and frequency $\omega_k$, $\mathbf{f}_k$ being the profile of the fundamental transverse mode of the waveguide at wavenumber $k$, so that $g_k(\mathbf{r})=\sqrt{\omega_k/(2\epsilon_0\hbar)}\mathbf{d}\cdot \mathbf{f}_k e^{ikz}/\sqrt{L}$.  Assuming that the relevant $k$ wavenumbers form a relatively narrow band around $k_0=\omega_0/c$, we take a linear dispersion relation $\omega_k=\omega_0+kc$ and ignore the dependence of the transverse profile on $k$, $\mathbf{f}_k(x,y)=\mathbf{e}_0 f_0(x,y)$.

The atoms are trapped along the waveguide between $z=0$ and $z=L$ with a constant density (per unit length) along $z$, $n_a$, and a transverse profile $p(x,y)$ (per unit area) of width $w$ around some point $r_a=(x_a,y_a)$. Considering the trapping width $w$ to be much smaller than the spatial variation of $f_0(x,y)$ (e.g. $\sim60$nm and $250$nm, respectively, in \cite{RAU1}), we can take $\rho(\mathbf{r})=n_a p(x,y)\approx n_a \delta(x-x_a)\delta(y-y_a)$. Then, identifying the effective area $f_0(x_a,y_a)=1/\sqrt{A}$ and the collective atomic operators $\hat{\sigma}_{ij}(z)=\int dx \int dy p(x,y) \hat{\sigma}_{ij}(\mathbf{r})$, and moving also to an interaction picture with respect to $H_A+\sum_k\hbar\omega_0\hat{a}_k^{\dag}\hat{a}_k$, we obtain the effectively 1d Hamiltonian, Eq. (\ref{HDD}) in the main text.
\subsection{Weak probe approximation}
Using the Hamiltonian (\ref{HDD}), we derive the following Heisenberg equations of motion,
\begin{eqnarray}
&&\left(\partial_t+c\partial_z\right)\hat{\mathcal{E}}(z)=n_a g^{\ast} \tilde{\sigma}_{ge}(z),
\nonumber \\
&&\partial_t\tilde{\sigma}_{ge}(z)=-\Omega^{-i\delta_c t}\tilde{\sigma}_{gd}(z)+g\hat{\mathcal{E}}(z)\left[\hat{\sigma}_{ee}(z)-\hat{\sigma}_{gg}(z)\right]
\nonumber\\
&&\quad\quad\quad\quad\quad-\gamma\tilde{\sigma}_{ge}+\hat{F},
\nonumber \\
&&\partial_t\tilde{\sigma}_{gd}(z)=\Omega^{\ast}e^{i\delta_c t}\tilde{\sigma}_{ge}(z)+g\hat{\mathcal{E}}(z)\hat{\sigma}_{ed}(z)e^{ik_cz}
\nonumber\\
&&\quad\quad\quad\quad\quad-i n_a \int dz'U(z-z')\hat{\sigma}_{dd}(z')\tilde{\sigma}_{gd}(z),
\label{EOM1}
\end{eqnarray}
with $\tilde{\sigma}_{ge}=\hat{\sigma}_{ge}e^{-ik_0z}$ and $\tilde{\sigma}_{gd}=\hat{\sigma}_{ge}e^{i(k_c-k_0)z}$. The second line of the equation for $\partial_t\tilde{\sigma}_{ge}$ describes the effect of the reservoir formed by all photonic modes not included in the Hamiltonian (non-guided modes), namely, spontaneous emission from the level $|e\rangle$ and its corresponding Langevin noise operator.
We now assume that all atoms are initially in the state $|g\rangle$ and that the probe field $\hat{\mathcal{E}}$ is weak enough so that the photon density in the medium is much smaller than the atom density hence the \emph{individual} atoms respond linearly to the probe (i.e. the transition $|g\rangle\rightarrow|e\rangle$ is far from saturation): $\hat{\sigma}_{ij}=O(\hat{\mathcal{E}})$ for $i\neq j$,  and $\hat{\sigma}_{gg}\sim 1$. Applying such linearization to Eqs. (\ref{EOM1}) while retaining the nonlinearity due to the interaction $U$ between \emph{different} atoms, we can eliminate $\tilde{\sigma}_{ge}$ and obtain the coupled equations for $\tilde{\sigma}_{gd}$ and $\hat{\mathcal{E}}$, Eqs. (\ref{EOM}) from the  main text.
\subsection{Adiabatic expansion}
The equations of motion for the polaritons, Eqs. (\ref{EOMpol}), were obtained directly from the ones for the atoms and weak probe field, Eq. (\ref{EOM}) without further approximations. In  order to arrive at our central equation, Eq. (\ref{NLSE}), we then perform an adiabatic expansion in the spirit of Ref. \cite{FL}. The main idea is to consider an overall EIT configuration which is very close to both single-photon and two-photon resonances, namely, a slowly varying probe envelope $\hat{\mathcal{E}}(z)$ and small detunings for the coupling field. In turn, this implies small temporal derivatives of both polariton fields on the one hand, and small linear and nonlinear detunings on the other hand, respectively. Namely, our small parameter in the expansion, denoted $T^{-1}$, scales as
\begin{equation}
T^{-1} \gtrsim \partial_t \hat{\Psi}, \partial_t \hat{\Phi}, \delta_c,\hat{S}.
\label{T}
\end{equation}
For example, in the adiabatic limit $T^{-1}\rightarrow 0$ we take the lowest order in Eqs. (\ref{EOMpol}), obtaining
\begin{equation}
(\partial_t+c\cos^2\theta\partial_z)\hat{\Psi}=-i\sin^2\theta[\delta_c+\hat{S}(z)]\hat{\Psi}, \quad\hat{\Phi}=0.
\label{a0}
\end{equation}
In order to find the lowest order corrections to this limit, we first go back to Eqs. (\ref{EOMpol}) and insert the equation for $\hat{\Phi}$ into that of $\partial_t\hat{\Psi}$, obtaining an equation of the form
\begin{equation}
(\partial_t+c\cos^2\theta\partial_z)\hat{\Psi}=D'+D'',
\label{a1}
\end{equation}
where $D'$ are terms associated with dispersion effects and $D''$ with dissipative effects. In the following we focus on each of these terms separately.
\\\\
\emph{Dissipative terms}. We find,
\begin{eqnarray}
D''&=&-\frac{\sin^2\theta\cos^2\theta}{|\Omega|^2}\gamma[c\partial_z-i(\delta_c+\hat{S})][\partial_t+i(\delta_c+\hat{S})]\hat{\Psi}
\nonumber\\
&&+\frac{\sin\theta\cos^3\theta}{|\Omega|^2}\gamma[c\partial_z-i(\delta_c+\hat{S})][\partial_t+i(\delta_c+\hat{S})]\hat{\Phi}
\nonumber\\
&&-\frac{\sin\theta\cos^2\theta}{\Omega}\sqrt{n_a}[c\partial_z-i(\delta_c+\hat{S})]\hat{F}.
\label{D2}
\end{eqnarray}
For the lowest order correction to the adiabatic limit, we insert $\hat{\Psi}$ and $\hat{\Phi}$ from the adiabatic solution, Eqs. (\ref{a0}) into $D''$. Then, the second line in Eq. (\ref{D2}) vanishes, whereas the first line scales as $\sin^2\theta(\gamma/|\Omega|^2)T^{-2}\hat{\Psi}$ [noting from Eqs. (\ref{a0}) and (\ref{T}) that $c\cos^2\theta\partial_z\sim T^{-1}$]. Then, this term of $D''$ is negligible with respect to the leading contribution of the left hand side of Eq. (\ref{a1}) (scaling as $T^{-1}\hat{\Psi}$) if
\begin{equation}
T^{-1}\ll \frac{|\Omega|^2}{\gamma}\frac{1}{\sin^2\theta}.
\label{cond}
\end{equation}
For typical EIT conditions with $v\ll c$, we have $\sin^2\theta\approx 1$ and we get that the single- and two-photon detunings of our system, $T^{-1}$, should be within the single-atom EIT transparency bandwidth $|\Omega|^2/\gamma$. It should be noted however that the usefulness of the single-atom transparency window in predicting the losses in our system relies on the expansion in $T^{-1}$, i.e. on very narrow detunings, and that in the Autler-Townes regime for example, considerable losses are expected for $T^{-1}\sim \Omega\ll|\Omega|^2/\gamma$. Nevertheless, the real condition for lossless propagation has to take into account the entire medium and not just a single atom. It can be found by noting that the first line in $D''$ provides a decay rate for $\hat{\Psi}$ from which we find that the losses in a medium of length $L$ are negligible for
\begin{equation}
T^{-1}\ll|\frac{\Omega|^2}{\gamma\sqrt{OD}}\equiv\delta_{tr},
\label{cond2}
\end{equation}
where $\delta_{tr}$ is the transparency window of the entire medium, as usual \cite{FL}, which for large $OD$ can become much narrower than both $\Omega|^2/\gamma$ and $\Omega$.

Finally, the quantum fluctuations induced by the reservoir formed by the vacuum of all non-guided modes, captured by the $\hat{F}$ term, are related to the losses by the fluctuation-dissipation theorem. Therefore, when losses are negligible, so is the effect of these fluctuations, upon comparison to the reservoir-free quantum fluctuations of $\hat{\Psi}$. We have verified this, using the equation for $\hat{\Phi}$ in Eq. (\ref{EOMpol}), by showing that the variance which $\hat{F}$ adds to $\hat{\Phi}$ is negligible with respect to the vacuum noise of $\hat{\Phi}$.
\\\\
\emph{Dispersive terms}. We find,
\begin{eqnarray}
D'&=&-\frac{\sin^2\theta\cos^2\theta}{|\Omega|^2}[c\partial_z-i(\delta_c+\hat{S})]\partial_t[\partial_t+i(\delta_c+\hat{S})]\hat{\Psi}
\nonumber\\
&&+\frac{\sin\theta\cos^3\theta}{|\Omega|^2}[c\partial_z-i(\delta_c+\hat{S})]\partial_t[\partial_t+i(\delta_c+\hat{S})]\hat{\Phi}
\nonumber\\
&&-i\sin^2\theta(\delta_c+\hat{S})\hat{\Psi}.
\label{D1}
\end{eqnarray}
Inserting the adiabatic solutions, Eqs. (\ref{a0}), into the above expression, we obtain $D'$ up to order $T^{-3}\hat{\Psi}$. Then, for approximately lossless propagation as in Eq. (\ref{cond}), we take $D''\approx 0$, and Eq. (\ref{a1}) takes the form
\begin{eqnarray}
(\partial_t+c\cos^2\theta\partial_z)\hat{\Psi}=(i\hat{A}+\hat{B}\partial_z+i\hat{C}\partial^2_z+\hat{D}\partial^3_z)\hat{\Psi},
\label{a3}
\end{eqnarray}
with hermitian (real) coefficients $\hat{A},\hat{B},\hat{C}$ and $\hat{D}$ expanded up to order $T^{-3},T^{-2},T^{-1}$ and $T^0$, respectively. Taking the lowest non-vanishing order for each coefficient, we find $\hat{S}\approx \hat{\delta}_{NL}$ and end up with Eq. (\ref{NLSE}) from the main text, plus an extra $\hat{D}$-coefficient term with $\hat{D}=-(\sin^2\theta/|\Omega|^2)v^3$. Nevertheless, for the regime explored in this work, namely, that of a probe-pulse bandwidth smaller than the combined detuning $\delta_c+\hat{\delta}_{NL}$, such cubic dispersion term is negligible (this was also verified by repeating calculations while including this term).

Finally, it should be noted that our derivation of Eq. (\ref{NLSE}) can be shown to be consistent with a similar adiabatic expansion in Ref. \cite{FL}, upon relating the linear and nonlinear detunings $\delta_c+\hat{S}$ in Eqs. (\ref{EOMpol}) to the time-dependence of the EIT mixing angle in \cite{FL} (Sec. V therein).

\subsection{Significant dispersion in the lossless regime}
The question is now under which conditions the above \emph{dispersive} corrections to adiabaticity, giving rise to the $\partial^2_z$ term in Eq. (\ref{NLSE}), are significant while losses are nevertheless negligible. By comparing the first line of $D'$ to that of $D''$ in Eqs. (\ref{D1}) and (\ref{D2}), respectively, we observe that dispersive effects are dominant over dissipative ones when $T^{-1}\gg \gamma$. On the other hand, for lossless propagation we have from (\ref{cond}), $T\gg |\Omega|^2/\gamma$, leading to the Autler-Townes condition $\Omega\gg\gamma$, as also explained In sec. II A and Fig. 2. More realistically, we consider lossless propagation with the condition (\ref{cond2}), leading to the requirement
\begin{equation}
\gamma\ll T^{-1}\ll\frac{|\Omega|^2}{\gamma\sqrt{OD}},
\label{disp}
\end{equation}
and hence to a modified Autler-Townes-like condition $\gamma OD^{1/4}\ll \Omega$. In this work, we consider a case where deviations from adiabaticity are mainly due to the overall detuning of the coupling field, $\Delta_c=\delta_c+\delta_{NL}$ (in analogy to the intuitive explanation of sec. II A), so that the condition (\ref{disp}) has to be satisfied for $T^{-1}\sim \Delta_c$.

\section{Bogoliubov theory for a nonlocal interaction}
\subsection{Excitation spectrum}
The polariton wave excitation spectrum (dispersion relation) around the CW solution, Eq. (\ref{bog}), is found as follows. We first write the polariton field as a sum of an average CW $\langle\hat{\Psi}(z,t)\rangle=\psi(t)$ and a small fluctuation $\hat{\varphi}(z,t)$,
\begin{equation}
\hat{\Psi}(z,t)=\psi(t)+\hat{\varphi}(z,t),
\label{1}
\end{equation}
where the average CW solution is that of self-phase modulation, $\psi(t)=\psi_0e^{-i(\alpha \delta_c+n_pU_0)t}$ with $\psi_0=\sqrt{n_p/\alpha^2}e^{i\phi}$, $U_0=U_{k=0}$, $U_k$ being the spatial Fourier transform of the potential $U(z)$,
\begin{equation}
U_k=\int_{-\infty}^{\infty} dz U(z) e^{-ikz}.
\label{Uk}
\end{equation}
The integral for $U_0$ that arises in the above CW solution $\psi(t)$ runs over $z'=0$ to $z'=L$ around some point $z$, $\int_{0}^{L} dz' U(z-z')$, and not from $z'=-\infty$ to $z'=\infty$ around $z=0$ as in the definition (\ref{Uk}). However, for a symmetric potential [$U(z)=U(-z)$] with range $l$, and points $z$ well within the medium, i.e. $l<z<L-l$, the edges $z=0$ and $z=L$ have no effect, so that $U_0$ can be  approximated as $\int_{0}^{L} dz' U(z-z')\approx \int_{-\infty}^{\infty} dz' U(-z')$ for all such points $z$. Thus, we may neglect the edge effects of a sufficiently long medium $L\gg l$.

Inserting Eq. (\ref{1}) into the nonlinear equation (\ref{NLSE}) in the main text and keeping terms only to linear order in the perturbation $\hat{\varphi}$, we obtain
\begin{eqnarray}
&&(\partial_t+v\partial_z)\hat{\varphi}(z)=-i(\alpha\delta_c+n_p U_0) \hat{\varphi}(z)
\nonumber\\
&&+i(n_p U_0/\alpha+\delta_c)Cv^2\partial_z^2\hat{\varphi}(z)-in_p\int_L dz'U(z-z')\hat{\varphi}(z')
\nonumber\\
&&-in_pe^{i2[\phi-(\alpha\delta_c+n_p U_0)t]}\int_Ldz'U(z-z')\hat{\varphi}^{\dag}(z').
\label{linNLSE}
\end{eqnarray}
This linearized equation is valid as long as the fluctuations $\hat{\varphi}$ around the mean $\psi$ are small, i.e. for
\begin{equation}
|\psi| \gg |\langle\hat{\varphi}\rangle|, \sqrt{\langle\hat{\varphi}^{\dag}\hat{\varphi}\rangle},\sqrt{\langle\hat{\varphi}^2\rangle}.
\label{cv}
\end{equation}
In order to find the dispersion relation of the fluctuations it is enough to consider the case of a classical field $\hat{\varphi}\rightarrow \varphi$. We then insert the ansatz for $\varphi(z,t)$ from Eq. (\ref{an}) into the linearized equation (\ref{linNLSE}), obtaining coupled equations for $u_k$ and $v_k$
\begin{equation}
\left(
  \begin{array}{cc}
    \omega_k^0+n_pU_k-\omega_k & -n_pU_k \\
    -n_pU_k & \omega_k^0+n_pU_k+\omega_k \\
  \end{array}
\right)
\left(
  \begin{array}{c}
    u_k \\
   v_k \\
  \end{array}
\right)
=0
\label{uv}.
\end{equation}
In order to arrive at the equations (\ref{uv}), we used
\begin{equation}
\int_0^L dz'U(z-z')e^{ikz'}\approx \int_{-\infty}^{\infty}U(\xi)e^{ik\xi}e^{ikz}= U_{k}e^{ikz},
\label{edge}
\end{equation}
which again amounts to neglecting edge effects, and where in the last equality we assumed that $U(z)$ is symmetric and real so that $U_{-k}=U_k$.
Eq. (\ref{uv}) has a nontrivial solution only if the determinant of the matrix on the left-hand side is zero, yielding an extra equation whose solution is the spectrum $\omega_k$ from Eq. (\ref{bog}).

\subsection{Dynamic Bogoliubov theory}
Here we address the quantum description of the dynamics of the fluctuations $\hat{\varphi}$ around the CW solution and arrive at the dynamical Bogoliubov transformation with the coefficients of Eq. (\ref{bog2}) in the main text. We expand the quantum field $\hat{\varphi}(z,t)$ in spatial Fourier modes $\hat{a}_k(t)$, $\hat{\varphi}(z)=\sum_k(1/\sqrt{L})e^{ikz}\hat{a}_k$, and the transformed modes $\hat{c}_k(t)$ as,
\begin{eqnarray}
\hat{\varphi}(z,t)&=&\sum_k\frac{1}{\sqrt{L}}e^{ik(z-vt)}e^{i[\phi-(\alpha\delta_c+n_pU_0)t]}\hat{c}_k(t),
\nonumber\\
\hat{a}_k(t)&=&e^{i\phi}e^{-i(\alpha\delta_c+n_pU_0+kv)}\hat{c}_k(t),
\label{c}
\end{eqnarray}
where the operators $\hat{a}_k(t)$ satisfy the equal-time commutation relations $[\hat{a}_k(t),\hat{a}^{\dag}_{k'}(t)]=\delta_{kk'}$ and hence so do the operators $\hat{c}_k(t)$.
Inserting Eq. (\ref{c}) into the linearized equation (\ref{linNLSE}) and neglecting edge effects as in Eq. (\ref{edge}), we obtain an equation for $\partial_t\hat{c}_k$. Then, upon taking its Hermitian conjugate, we end up with coupled equations of motion for $\hat{c}_k$ and $\hat{c}_{-k}^{\dag}$,
\begin{equation}
\partial_t
\left(
  \begin{array}{c}
    \hat{c}_k \\
   \hat{c}^{\dag}_{-k} \\
  \end{array}
\right)
=
-i\left(
  \begin{array}{cc}
    \omega_k^0+n_pU_k & n_pU_k \\
    -n_pU_k & -\omega_k^0-n_pU_k \\
  \end{array}
  \right)
\left(
  \begin{array}{c}
    \hat{c}_k \\
   \hat{c}^{\dag}_{-k} \\
  \end{array}
\right).
\label{ck}
\end{equation}
By diagonalizing the matrix, we find the solution for $\hat{c}_k(t)$ from which we obtain the dynamics of $\hat{a}_k(t)$ as the Bogoliubov transformation from Eq. (\ref{bog2}) in the main text.

\subsection{Intensity correlations}
Intensity correlations are characterized by the normalized second order coherence function defined by \cite{MW}
\begin{equation}
g^{(2)}(z,z')=\frac{G^{(2)}(z,z')}{G^{(1)}(z)G^{(1)}(z')}
\label{g2def}
\end{equation}
with
\begin{eqnarray}
G^{(2)}(z,z')&=&\langle \hat{\Psi}^{\dag}(z,t)\hat{\Psi}^{\dag}(z',t)\hat{\Psi}(z',t)\hat{\Psi}(z,t)\rangle,
\nonumber\\
G^{(1)}(z)&=&\langle\hat{\Psi}^{\dag}(z,t)\hat{\Psi}(z,t)\rangle.
\label{G}
\end{eqnarray}
Here we assume that all points of the field $\hat{\Psi}(z)$ have experienced the same duration of interaction $t$ upon arrival at the detector. The detector measures at different times different points of the field $\hat{\Psi}(z)$, so that the correlations between detector-signals at different times are in fact correlations of the field in space as in $g^{(2)}(z,z')$ which quantifies the autocorrelation of the field between $z$ and $z'$.

Upon expanding the field in its longitudinal Fourier modes, we recall that the $k=0$ mode can be approximated by the strong average CW solution,
\begin{eqnarray}
&&\hat{\Psi}(z,t)=\frac{1}{\sqrt{L}}\sum_ke^{ikz}\hat{a}_k(t)\approx \frac{a_0(t)}{\sqrt{L}}+\frac{1}{\sqrt{L}}\sum_{k\neq0}e^{ikz}\hat{a}_k(t),
\nonumber\\
&&a_0(t)/\sqrt{L}=\sqrt{n_p/\alpha^2}e^{i\phi}e^{-i(\alpha \delta_c+n_pU_0)t},
\label{psi}
\end{eqnarray}
where $\hat{a}_k(t)$ is given by Eq. (\ref{bog2}) in the main text. Inserting Eq. (\ref{psi}) into $G^{(1)}(z)$ from Eq. (\ref{G}), we find
\begin{equation}
G^{(1)}(z)=\frac{n_p}{\alpha^2}+\frac{1}{L}\sum_{k\neq 0}\left[|\mu_k|^2N_k+|\nu_k|^2(N_k+1)\right],
\label{G1}
\end{equation}
where we have assumed the following statistics of the initial fluctuations $\hat{a}_{k\neq 0}$: $\langle\hat{a}_{k}(0)\rangle=0$, $\langle \hat{a}_k^{\dag}(0)\hat{a}_{k'}(0)\rangle=N_k\delta_{kk'}$ with $N_k=N_{-k}$, and $\langle \hat{a}_k(0)\hat{a}_{k'}(0)\rangle=0$.
Moving to the second-order coherence, we insert Eq. (\ref{psi}) into $G^{(2)}$ from Eq. (\ref{G}), keeping terms only to second order in the fluctuations $\hat{a}_{k\neq 0}$, in accordance with the assumptions (\ref{cv}), and obtain
\begin{eqnarray}
&&G^{(2)}(z,z')=\frac{n_p^2}{\alpha^4}+2\frac{n_p}{\alpha^2}
\nonumber\\
&&\times\frac{1}{L}\sum_{k\neq 0}\left\{
\left[|\mu_k|^2N_k+|\nu_k|^2(N_k+1)\right]\left(1+\cos[k(z-z')]\right)\right.
\nonumber\\
&&\left.+|\nu_k||\mu_k|(1+2N_k)\cos[k(z-z')-\phi_k]\right\},
\label{G2}
\end{eqnarray}
where $\phi_k=\mathrm{arg}(\mu_k\nu_k)$. Inserting $G^{(1)}$ and $G^{(2)}$ from Eqs. (\ref{G1}) and (\ref{G2}) in Eq. (\ref{g2def}), we note that to lowest order in the fluctuations we can take $G^{(1)}\approx n_p/\alpha^2$ such that $g^{(2)}\propto G^{(2)}$. Then, upon taking the continuum limit $(1/L)\sum_k\rightarrow(1/2\pi)\int_{-\infty}^{\infty}dk$, we split the integral into its positive and negative $k$-values and obtain Eq. (\ref{g2}) from the main text.

\section{Laser-induced dipole-dipole interaction}
\label{ALIDDI}
The laser-induced interaction potential between a pair of atoms, e.g. at positions $z_1$ and $z_2$ along the $z$-axis, is given by \cite{LIDDI}
\begin{equation}
U(z_1,z_2)=-\frac{|\Omega_L|^2}{\delta_L}\hbar \Delta_{dd}(z_1,z_2)\cos[k^z_L(z_1-z_2)],
\label{LIDDI}
\end{equation}
with the same notations from Sec. V (and Fig. 1c) and where $\Delta_{dd}(z_1,z_2)$ is the so-called resonant dipole-dipole interaction evaluated at the laser frequency $\omega_L$. $\Delta_{dd}(z_1,z_2)$ is responsible for dispersive interactions (excitation exchange) between the atoms mediated by virtual excitations of the photon modes that couple to the atoms. Therefore, the spatial dependence of $\Delta_{dd}(z_1,z_2)$, which in turn determines the potential $U(z_1,z_2)$, directly follows that of the photon modes. And in a confined geometry, these can alter dramatically and lead to modified and long range dipolar interactions $\Delta_{dd}$ and potentials $U$ \cite{RDDI,LIDDIna,CHA,KUR,LIDDI}. The case considered here, of a dipolar interaction at a frequency inside the bandgap of waveguide grating, leads to $\Delta_{dd}(z_1,z_2)\propto \eta\cos[k_B(z_1-z_2)]e^{-|z_1-z_2|/l}$, with $\eta,l\propto1/\sqrt{\omega_u-\omega_L}$, where $\omega_u$ is the frequency of the upper bandedge \cite{LIDDIna,CHA,RDDI}. Therefore, for $\omega_L$ not too far from the bandedge, strong and long-range interaction potentials can be achieved.

\section{Quantitative illustration}
\label{AQUA}
Let us specify the system parameters used in the illustration in Figs. 3 and 4. Motivated by the experiment in Ref. \cite{RAU1}, we take the wavelength, dipolar matrix element and radiative decay rate of the  D1 line of Cs atoms \cite{Cs} as the typical parameters of all dipolar atomic transitions in Fig. 1, and consider atoms trapped at a distance $r_a\sim 500$ nm from the center of a tapered fiber with radius $a=250$ nm and refractive index  $n=1.452$, leading to an effective area of the fiber's HE$_{11}$ transverse mode $A=4.42$ $\mu$m$^2$ \cite{LIDDIna}. Taking an atom density of $n_a\sim16\times10^6$ m$^{-1}$, length $L=2.68$ cm, and coupling laser with Rabi frequency $\Omega=4\times10^8$ s$^{-1}$ ($2\times10^8$ s$^{-1}$ and $1.2\times10^8$ s$^{-1}$ in Figs. 3b and 4, respectively) and detuning $\delta_c=-3.84\times10^7$ s$^{-1}$ in Fig. 3a ($4.795\times10^6$ s$^{-1}$ and $-3.64\times10^6$ s$^{-1}$ and in Figs. 3b and 4), we obtain $\alpha=0.999839$ ($0.99996$ and $0.999986$ in Figs. 3b and 4), $v=48216$ ms$^{-1}$ ($12055$ ms$^{-1}$ in $4340$ ms$^{-1}$ in Figs. 3b and 4, respectively) and $\delta_{tr}=8.66\times10^7$ s$^{-1}$ ($2.16\times10^7$ s$^{-1}$ and $7.79\times10^6$ s$^{-1}$ in Figs. 3b and 4), so that Figs. 3a,b and 4a,b are effectively cut at $k=q_{tr}=\delta_{tr}/v=1795$ m$^{-1}$. For the Bragg grating imprinted on the fiber/waveguide \cite{KIM2,HAK}, we assume periodic perturbations $\Delta n=0.02$ of the refractive index about $n$ with period of length $\Lambda=396$ nm.
For the laser induced interaction we take a detuning $\delta_L=-2\pi\times0.65$ GHz and intensity $I=2\times10^4$ Wm$^{-2}$ ($0.5\times10^4$ Wm$^{-2}$ and $10^2$ Wm$^{-2}$ in Figs. 3b and 4), yielding $U_L=2\eta R_{fs}=1.14\times10^6$ s$^{-1}$ ($2.85\times10^5$ s$^{-1}$ and $5696$ s$^{-1}$ in Figs. 3b and 4), where $\eta\sim 12$ is the ratio between emission to the fiber grating modes and to free-space, and $R_{fs}=\gamma|\Omega_L|^2/(2\delta_L^2)=47443$ s$^{-1}$ ($11860$ s$^{-1}$ and $237$ s$^{-1}$ in Figs. 3b and 4) is the resulting scattering rate to free space form the $|d\rangle\rightarrow|s\rangle$ transition, leading to a potential range of $l\approx3027\lambda_L\approx0.1L\approx0.0027$m. The orientation angle is taken to be $\theta_L=0.131$, so that $k_R=k_L\cos\theta_L-k_B=1019$ m$^{-1}$. We assume a CW background with a power $2\times10^{-12}$ W in Fig. 3a ($10^{-10}$ W and $4\times10^{-10}$ W in Figs. 3b and 4) , giving $n_p\approx187$ m$^{-1}$ ($37347$ m$^{-1}$ and $414981$ m$^{-1}$ in Figs. 3b and 4). For the $N_k(t)$ and $g^{(2)}$ calculations in Figs. 4b,c, we consider an initial Gaussian spectrum of intensity fluctuations, $N_k=N_0e^{-(k/q_{tr})^2}$, with $N_0=5$. Finally, wherever a comparison with the local-interaction case is made, the local potential is taken as $U(z)=(1/8)U_L l\delta(z)$.

\section{Numerical simulations of Eq. (\ref{NLSE})}
\label{ANUM}
In order to obtain the $\omega_k$ spectrum, we perform numerical simulations of the full nonlinear equation (\ref{NLSE}). We use a three-term splitting method, capable of dealing with the nonlocal and nonlinear dispersion coefficient that appears in Eq. (\ref{NLSE}), unlike the more common (two-term) split-step method. We study the dynamics of the field, initially comprised of the CW solution $\psi$ and weak perturbations $\varphi$ at a spatial frequency $k$, and extract the field's temporal oscillation frequency $\omega(k)$ leading to the spectrum $\omega_k$. For weak enough perturbations where the Bogoliubov theory applies, the results of the simulation of  Eq. (\ref{NLSE}) agree very well with those of the simpler split-step method simulations, where the nonlinear dispersion term  $\hat{\delta}_{NL}Cv^2\partial_z^2$ is approximated by its mean-field value, $\delta_{NL}= n_pU_0/\alpha$, so that the equation becomes of semilinear type.
For the case of instability (Fig. 4b,c), we begin with the initial intensity fluctuations with a spectrum $N_k$ and run split-step simulations with the mean-field nonlinear dispersion term $\delta_{NL}= n_pU_0/\alpha$. Then, after propagation time $t$, we measure $N_k(t)=\langle a_k^{\ast}(t)a_k(t)\rangle$ and $g^{(2)}=\langle I(z)I(z')\rangle/(\langle I(z)\rangle\langle I(z')\rangle)$, with $I(z)=|\Psi(z)|^2$ for a classical field $\Psi$.

\section{Scattering and imperfections}
\label{ASC}
In the following we address the main loss and dephasing mechanisms which affect the polariton evolution. For each of them we estimate the rate at which they decohere the polariton field; then, by comparing the analytical expression of this decoherence rate with that of the spectra $\omega_k$ (or $\gamma_k=\mathrm{Im}\omega_k$ in the case of instability), we study how the limitations it imposes on the observability of the spectra can be relaxed.

\subsection{Scattering of off-resonant laser photons}
Beginning with the illumination of the $|d\rangle\rightarrow|s\rangle$ atomic transition by the off-resonant laser $\Omega_L$, it leads to an interaction potential $U(z)$, but is also accompanied by scattering of photons to non-guided modes at a rate $R_{fs}=\gamma|\Omega_L|^2/(2\delta_L^2)$ \cite{LIDDI}, which decoheres the spin wave $\bar{\sigma}_{gd}$, and hence the polariotn, at rate $R_{fs}$. This rate has to be compared to the frequency scale $U_L=2\eta R_{fs}$ of the coherent interaction-related effects such as $\omega_L$, where $\eta$ is a geometrical factor related to the mode-density of the guided modes. For the case of waveguide-grating modes, $\eta$ can be larger than 1 ($\eta\sim12$ in our numerical example, see Appendix \ref{AQUA}) so coherent effects prevail \cite{LIDDIna}. In fact, a deeper reason for the dominance of the coherent effects discussed here (the spectra $\omega_k$ and $\gamma_k$) results from the fact that these are all \emph{cooperative} effects, proportional to the density of excitations (EIT polaritons) $n_p$ in the system, whereas the scattering $R_{fs}$ is essentially a \emph{single-atom} effect since photons are scattered to non-guided, non-confined modes (which mediate much weaker cooperative effects) \cite{LIDDIna}. Therefore, $\omega_k$ scales as $n_p U_L\propto n_p \eta R_{fs}$ [Eq. (\ref{bog}))] so that even for small $\eta$, it can be made much larger than $R_{fs}$ by e.g. increasing $n_p$ (increasing the laser power of the background CW probe field).
Indeed, for the specific numerical example taken here (Appendix \ref{AQUA}), we obtain $R_{fs}=47443$ s$^{-1},11860$ s$^{-1},237$ s$^{-1}$ for the roton, anti-roton and instability cases, respectively, yielding decoherence rates $R_{fs}$ much smaller than the typical frequency resolution of the corresponding spectra $\omega_k$ of the roton and anti-roton (Figs. 3a and 3b), and $\gamma_k$ of the instability (Fig. 5a).

\subsection{EIT propagation losses}
Next, we recall that lossless propagation of EIT polaritons is only exact for CW polaritons and vanishing coupling-field detuning $\Delta_c=0$. As seen in Fig. 2, the EIT susceptibility in fact exhibits a finite spectral "transparency window", leading to a loss rate \cite{FIM}
\begin{eqnarray}
R_{EIT}&=&(1/2)k_0 v \mathrm{Im}\chi(\Delta_p,\Delta_c)
\nonumber\\
&=&v \frac{OD}{L}\frac{2(\Delta_p-\Delta_c)^2\gamma^2}{4\gamma^2(\Delta_p-\Delta_c)^2+[\Omega^2-4\Delta_p(\Delta_p-\Delta_c)]^2},
\nonumber\\
\label{EIT}
\end{eqnarray}
$\gamma$ being the width of the atomic level $|e\rangle$ and $OD$ the optical depth of the medium (see main text).  For simplicity, we evaluate $R_{EIT}$ at the center of the pulse ($\Delta_p=0$) and for $\Delta_c=\delta_c+\delta_{NL}$, where $\delta_{NL}=n_p U_0/\alpha$ is the mean-field value of the nonlinear detuning. Then, the effect of this EIT-related decoherence rate on $\omega_k$ can be made much smaller by the following: consider decreasing both $n_p$ and $\delta_c$ by a factor $f$, resulting in a decrease of $\omega_k$ by the same factor [see Eq. (\ref{bog})]. On the other hand, Eq. (\ref{EIT}) reveals that $R_{EIT}$ scales approximately as $\Delta_c^2=(\delta_c+n_p U_0/\alpha)^2$ and hence decreases by $f^2$, so that it can become much smaller than $\omega_k$ and its related effects (the scaling $R_{EIT}\sim\Delta_c^2$ holds for large $\Omega$ in correspondence with the Autler-Townes regime of EIT that we consider).

Referring back to our numerical example we obtain $R_{EIT}\sim0.7\times10^6$ s$^{-1}$ for the roton case, so that this decoherence becomes comparable to the frequency resolution of the main features in $\omega_k$ [e.g. the difference between the roton and usual (local-interaction) Bogoliubov spectra at the roton "dip" around $k_R$ is about $0.15\times10^6$ s$^{-1}$, see Fig. 3a]. Then, decreasing (multiplying) both $n_p$ and $\delta_c$ by a factor $f=0.1$ we obtain $R_{EIT}\sim0.007\times10^6$ s$^{-1}$ whereas the typical frequency resolution for the roton spectrum becomes $0.015\times10^6$ s$^{-1}$, following the expected $f^2$  and $f$ scalings, respectively, and yielding a much suppressed effect of the losses. For the anti-roton case, again with the numbers from Appendix \ref{AQUA}, we have $R_{EIT}\sim0.18\times10^5$ s$^{-1}$, which is already much smaller than the typical resolution of the anti-roton feature, $\sim1\times10^5$ s$^{-1}$ (Fig. 3b). Finally, in the case of instability we find $R_{fs}\approx0.86\times10^5$ s$^{-1}$, which is smaller than but still rather close to the peak value of the amplification rate $\gamma_k$ in Fig. 5a. This can be improved by again decreasing $n_p$ and $\delta_c$ by $f=0.1$ and obtaining $R_{fs}\approx0.0086\times10^5$ s$^{-1}$ with $\gamma_k$ and its features reduced only by $f=0.1$.

\begin{figure}
\begin{center}
\includegraphics[scale=0.24]{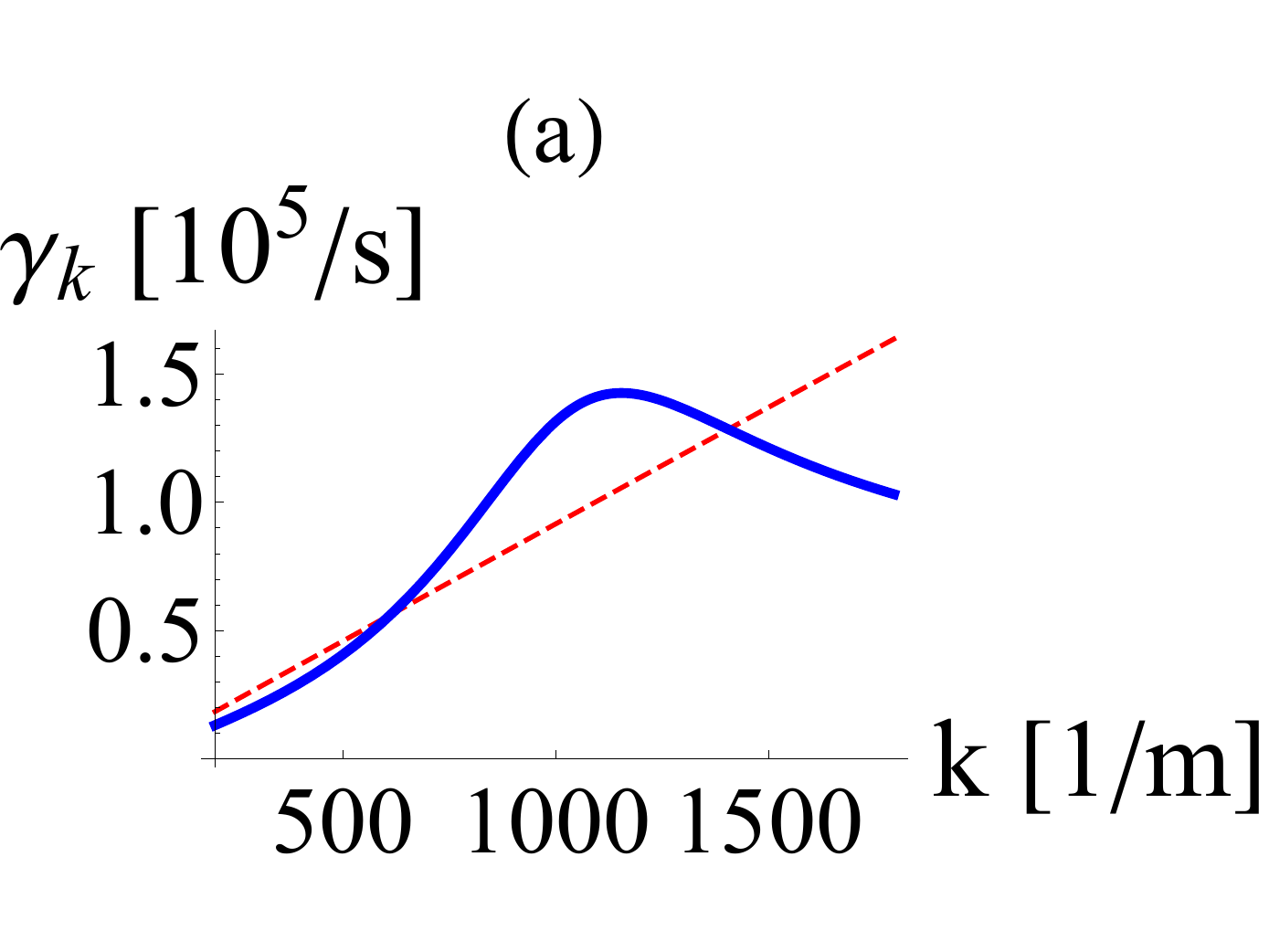}
\includegraphics[scale=0.24]{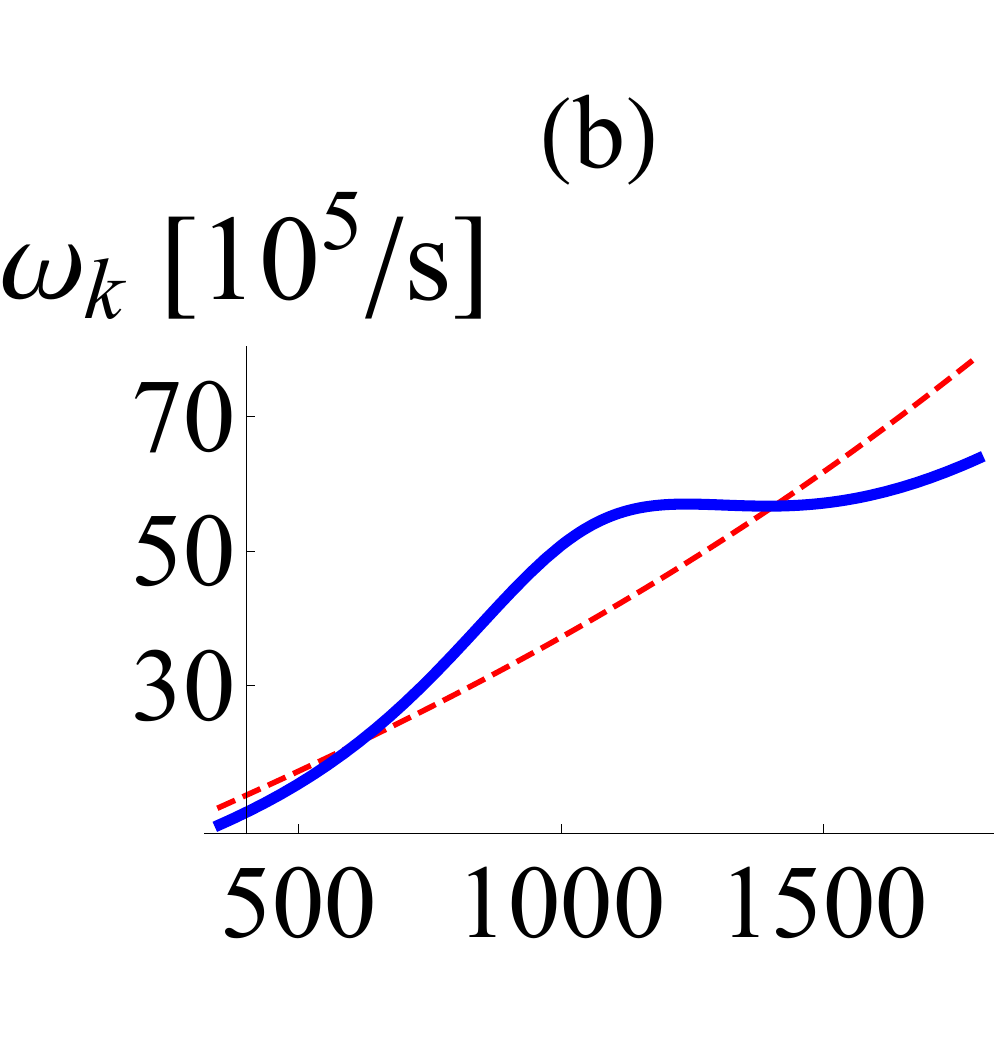}
\includegraphics[scale=0.24]{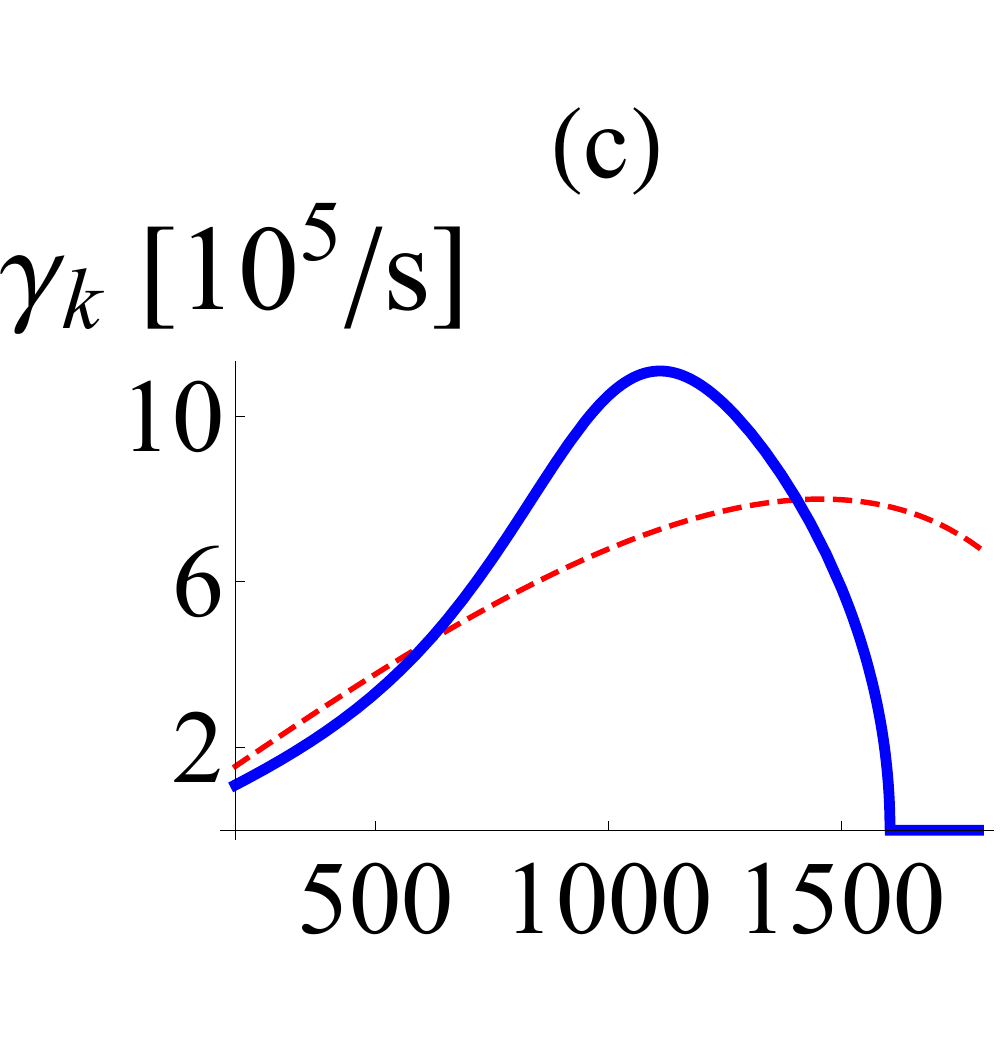}
\caption{\small{
(a) Instability case. Spectrum of amplification $\gamma_k=\mathrm{Im}\omega_k$ for the example considered in the main text (as in Fig. 4a therein): nonlocal potential (blue solid curve) and local potential (red dashed curve). (b) Overcoming cooperative decoherence $R^{(2)}_{im}$: anti-roton case. Same parameters as in the main text (Fig. 3b therein) apart from $\Omega$ which is increased by a factor 12. (c) Same as (b), however this time for the case of instability $\gamma_k$ and where $\Omega$ is increased by a factor 8.5.
}}
\end{center}
\end{figure}

\subsection{Scattering by imperfections in the waveguide structure}
The effect of imperfections on the guided modes $k$ is typically described by adding an imaginary part $-i\kappa$ to their mode frequencies. This can be consistently shown by e.g. considering pure modes which are dipole-coupled to scatterers modelled as lossy and localized harmonic oscillators far-detuned from the modes $k$ \cite{IMP}. Recalling that the interaction $U(z)$ between the atoms is mediated by the photon modes $k$, and taking their width $\kappa$ into account, $U$ now becomes complex,
\begin{eqnarray}
U(z)&=&U'(z)-iU''(z),
\nonumber\\
U'(z)&=&-\cos(k^z_L z)\frac{\Omega_L^2}{\delta_L^2}\int_0^{\infty}d\omega \frac{G(\omega,z)(\omega-\omega_L)}{(\kappa/2)^2+(\omega-\omega_L)^2},
\nonumber\\
U''(z)&=&-\cos(k^z_L z)\frac{\Omega_L^2}{\delta_L^2}\int_0^{\infty}d\omega\frac{G(\omega,z)\kappa/2}{ (\kappa/2)^2+(\omega-\omega_L)^2},
\nonumber\\
\label{cLIDDI}
\end{eqnarray}
where $G(\omega,z)$ (real function) is the autocorrelation spectrum of the photon-modes reservoir coupled to a pair of atoms $z$-apart \cite{RDDI,LIDDI,IMP,CHA}. The existence of the imaginary part $-iU''$ reflects the fact that now the modes $k$ not only mediate a dispersive interaction between the atoms ($U'$), but also mediate scattering of laser photons $\Omega_L$ to non-guided modes via the atoms and the imperfections in the waveguide.

For the modes $k$ of a waveguide with a bandgap, as in our example, we have $G(\omega,z)\propto1/\sqrt{\omega-\omega_u}$, with $\omega_u$ the frequency of the upper bandedge \cite{RDDI,LIDDIna}, so the integrals in Eq. (\ref{cLIDDI}) get most of their contribution around $\omega\sim\omega_u$. Then, as a simple estimation, we consider the lowest order in $\kappa$, so that
\begin{equation}
U'(z)\approx U(z), \quad U''(z)\approx\frac{\kappa}{2\delta_u}U(z), \quad \delta_u\equiv\omega_u-\omega_L,
\label{imp}
\end{equation}
with $U(z)$ from Eq. (\ref{UL}) in the main text.
We now turn to the estimation of three distinct decoherence rates associated with this imperfection-induced loss mechanism.
\\\\
\emph{Single-atom scattering.--}
The resulting decoherence rate on a single atom is obtained by taking $z\rightarrow0$ in Eq. (\ref{imp}), yielding $R^{(1)}_{im}\sim(1/4)|U_L|(\kappa/\delta_u)$. We note that this is a \emph{single-polariotn} decoherence rate and hence can be made much smaller than the cooperative effects represented by $\omega_k$, by increasing $n_p$ (as explained for the $R_{fs}$ case above).
\\\\
\emph{Cooperative scattering.--}
Atoms located at a distance $z$ apart can also scatter photons in a cooperative manner, leading to a mutual decoherence rate $U''(z)$. The resulting \emph{cooperative decoherence} per atom is then obtained by summing over interfering mutual scattering from all $|d\rangle$-populated atoms in the medium, $R^{(2)}_{im}\sim 0.5 n_p\int_0^L dz'(\kappa/\delta_u)U(z')\sim 0.5(\kappa/\delta_u)n_p U_0$, where $n_p$ is the density of polariton excitations in the medium (which is roughly the density of $|d\rangle$-populated atoms). We now need to compare $R^{(2)}_{im}\sim 0.5(\kappa/\delta_u)n_p U_0$ to $\omega_k=\sqrt{\omega_k^0(\omega_k^0+2n_pU_k)}$. It is easy to show that $\omega^0_k\propto Cv^2\propto \Omega^2$, so that near the peak of $\omega_k$, where $2n_p U_k$ is dominant over $\omega^0_k$, we have $\omega_k\propto \Omega$. On the other hand, $R^{(2)}_{im}$ is independent of $\Omega$, so that in principle $\omega_k$ can always be made dominant over $R^{(2)}_{im}$ by increasing $\Omega$.

For our numerical example we first need to estimate $\kappa$: current state-of-the-art nano-waveguide-based cavities can have an imperfection-induced $Q$-factor of up to $\sim4\times10^5$ for a cavity resonance at $\sim 800$ nm, so that $\kappa\sim 5\times10^9$ s$^{-1}$. Going back to our numerical example from the main text, we typically have $\delta_u\sim10^9$ s$^{-1}$, yielding $R^{(2)}_{im}\sim 2.5n_p U_0$. Then, for the anti-roton case, we find $R^{(2)}_{im}\sim 41\times10^5$ s$^{-1}$, much larger than the anti-roton peak of a few $10^5$ s$^{-1}$ (Fig. 3b). This can be improved by increasing $\Omega$ by e.g. a factor of $f=12$, yielding the pronounced anti-roton spectrum of Fig. 5b. Similarly, for the case of instability, we find $R^{(2)}_{im}\sim 9.3\times10^5$ s$^{-1}$, much larger than the amplification rate $\gamma_k$ which is peaked at a value of about $1.5\times10^5$ s$^{-1}$. Again, this can be remedied upon increasing $\Omega$ by a factor $f=8.5$ for example, resulting in the amplification profile of Fig. 5c. For the roton case, this method of increasing $\Omega$ is less efficient due to the sensitivity of the roton "dip" feature to changes in $\Omega$, hence $R^{(2)}_{im}$ may pose an important limitation on the its observability.
\\\\
\emph{Polariton propagation losses.--}
Eq. (\ref{pol}) defines the polariotn $\hat{\Psi}$ as a superposition of an atom (spin wave) and a propagating photon. The origin of both $R^{(1)}_{im}$ and $R^{(2)}_{im}$ discussed above are in its atomic component, whereas its photonic component, essentially comprised of the modes $k$, also decays at a rate $\kappa$ due to photon propagation in the presence of imperfections. This means that the polariton field suffers from an additional decoherence given by $\kappa$ times the fraction of its photon component $\cos^2\theta$ (likewise, $R^{(1,2)}_{im}$ should be multiplied by $\sin^2\theta$, however as typical of EIT, we have $\sin^2\theta\approx 1$). Nevertheless, unlike all other decoherence rates discussed here ($R_{fs},R_{EIT}$ and $R^{(1,2)}_{im}$), this loss mechanism is \emph{independent} of the existence of the interaction $U(z)$ (it does not depend on the laser $\Omega_L$), and hence can be first measured in the absence of $U$ and then corrected for in measurements where $U$ is switched on.


\begin{thebibliography}{}
\bibitem{IMP} We have taken a Hamiltonian including atoms, guided modes and a collection of local far-detuned harmonic-oscillators (scatterers), which are themselves represnted by an open system due to their interaction with the continuum of non-guided modes. We then found $\kappa$ and rederived the expressions for the dipole-dipole interaction between atoms mediated by "open" guided modes with spectral width $\kappa$, arriving at Eq. (\ref{cLIDDI}).
\end{thebibliography}
\end{document}